\PassOptionsToPackage{dvipsnames}{xcolor}
\documentclass[acmsmall, screen, authorversion]{acmart}

\setcopyright{cc}
\setcctype{by}
\acmJournal{TOSEM}
\acmYear{2026} \acmVolume{1} \acmNumber{1} \acmArticle{}
\acmMonth{1} \acmPrice{} \acmDOI{10.1145/3790101}

\usepackage{tcolorbox}
\usepackage{xspace}
\usepackage[inline]{enumitem}
\usepackage{subcaption}
\usepackage{bbding}
\usepackage{url}
\usepackage{tabularx}
\usepackage{circledsteps}
\usepackage{graphicx}
\usepackage{footmisc}
\usepackage{soul}

\usepackage{listings}
\definecolor{mygreen}{rgb}{0,0.6,0}
\definecolor{mymauve}{rgb}{0.58,0,0.82}
\lstdefinelanguage{JavaScript}{
  keywords={typeof, new, true, false, catch, function, return, null, catch, switch, var, if, in, while, do, else, case, break, const, let, async, await, for, of},
  ndkeywords={class, export, boolean, throw, implements, import, this, console, log},
  sensitive=false,
  comment=[l]{//},
  morecomment=[s]{/*}{*/},
  morestring=[b]",
}
\lstdefinelanguage{Rust}{
    keywords={fn, let, mut, pub, struct, enum, impl, trait, where, match, if, else, for, in, while, loop, break, continue, return, async, await, move, dyn, const, static, ref, use, mod, crate, super, self, Self, as, extern, type, true, false, Option, Some, None, Result, Ok, Err, Box, Vec, String, usize, i32, u32, f64, bool, char, str},
    ndkeywords={println!, vec!, macro_rules!, assert_eq!, panic!},
    sensitive=true,
    comment=[l]{//},
    morecomment=[s]{/*}{*/},
    morestring=[b]',
    morestring=[b]",
}
\usepackage[capitalise]{cleveref}
\usepackage{pifont}

\usepackage{multicol}

\usepackage{makecell}
\usepackage{multirow}
\newcolumntype{C}[1]{>{\centering\arraybackslash}p{#1}}

\usepackage{amsmath,bm,amsthm}
\theoremstyle{definition}

\usepackage{algorithm}
\usepackage{algpseudocode}
\algnewcommand\algorithmicforeach{\textbf{for each}}
\algdef{SE}[FUNCTION]{Function}{EndFunction}[2]{\algorithmicfunction\ \textnormal{#1}\ (#2)}{\algorithmicend\ \algorithmicfunction}
\algdef{S}[FOR]{ForEach}[1]{\algorithmicforeach\ #1\ \algorithmicdo}
\algrenewcommand\alglinenumber[1]{\footnotesize #1}
\algrenewcommand\algorithmicrequire{\small \textbf{input:}}
\algrenewcommand\algorithmicensure{\small \textbf{output:}}
\algrenewcommand\algorithmicfunction{\textbf{Function}}
\algtext*{EndFunction}
\algtext*{EndIf}
\algtext*{EndFor}
\algtext*{EndWhile}

\usepackage{fontawesome}
\usepackage{xcolor}
\usepackage{framed}
\usepackage{textcomp}

\usepackage{microtype}
\newcommand{\code}[1]{``\texttt{\small #1}''}

\newcounter{rq}
\tcbuselibrary{breakable}
\newenvironment{summary}{
	\begin{tcolorbox}[
			arc=2mm,
			boxrule=0.3pt,
			left=2pt,
			right=2pt,
			top=1pt,
			bottom=1pt,
			colframe=black,
            enlarge top by=2pt,
            enlarge bottom by=2pt,
            breakable=true,
		]
        \textbf{\faLightbulbO\ Message\,\refstepcounter{rq}\therq{}:}
		}{\end{tcolorbox}}

\newcommand{\sd}{$\mathcal{D}$}
\newcommand{\svp}{$\mathcal{P}$}
\newcommand{\svpp}{$\mathcal{PC}$}
\newcommand{\sdvp}{$\mathcal{D}\&\mathcal{P}$}
\newcommand{\sdvpp}{$\mathcal{D}\&\mathcal{PC}$}

\begin{document}

\title{Can Emulating Semantic Translation Help LLMs with Code Translation? A Study Based on Pseudocode}

\author{Songqiang Chen}
\email{i9s.chen@connect.ust.hk}

\author{Congying Xu}
\email{congying.xu@connect.ust.hk}

\author{Jingyi Chen}
\email{jchenix@connect.ust.hk}

\author{Jialun Cao}
\authornotemark[1]
\email{jialuncao@cse.ust.hk}

\author{Jiarong Wu}
\email{jwubf@connect.ust.hk}

\author{Shing-Chi Cheung}
\authornote{Corresponding authors.}
\email{scc@cse.ust.hk}

\affiliation{
    \institution{The Hong Kong University of Science and Technology}
    \city{Hong Kong}\country{China}
}
\affiliation{
    \institution{Guangzhou HKUST Fok Ying Tung Research Institute}
    \city{Guangzhou}\country{China}
}

\begin{abstract}
Although large language models (LLMs) show promising potential in code translation, they still struggle to generate accurate translations using the commonly adopted direct code-to-code translation approach, which converts an original program into the target programming language (PL) in a single step. Inspired by the success of incorporating intermediate steps to guide LLMs in resolving challenging tasks, in this study, we explore pseudocode-based code translation. This approach emulates human semantic translation by first interpreting the original program's intent and logic into pseudocode and then implementing it in the target PL. To understand the effectiveness of this underexplored approach, we present a systematic empirical study on pseudocode-based code translation, aiming to investigate its helpfulness in enhancing the direct translation approach, illuminate its effective usage, and identify its limitations. By comparing direct and pseudocode-based translation on 9,690 translation tasks across six PLs with five popular LLMs, we found that pseudocode-based translation can effectively complement direct translation, particularly when translating from flexible to rigid PLs and handling a low-training-resource PL. Based on the findings, we suggest combining the translation results of both approaches for test-based selection to leverage their complementary strengths. We also reveal the advantages of pseudocode-based translation in decoupling the code understanding and generation burden on complicated programs and mitigating distractions from PL-specific implementations in original programs, as well as its limitations due to incorrect, incomplete, or ambiguous pseudocode. Our study sheds light on the effective use of pseudocode-based translation and provides evidence to help enhance LLMs in code translation.
\end{abstract}

\keywords{Code Translation, Pseudocode, Semantic Translation, Large Language Model}

\ccsdesc[300]{General and reference~Empirical studies}
\ccsdesc[300]{Software and its engineering~Software maintenance tools}

\maketitle

\section{Introduction}

Code translation, also known as transpilation, refers to the automatic conversion of a program written in one programming language (PL) to another while preserving its behaviors \cite{fse13statisticalcodetranslation, TransCoder,icse24translationllmsurvey}. 
With the rapid evolution of PLs and the diverse requirements of software applications, code translation has gained significant attention in both research and industry due to its wide range of practical applications. For example, code translation facilitates the migration of legacy software systems built based on obsolete PLs to modern PLs~\cite{modernization-1, modernization-2, modernization-3}.
It also facilitates efficient prototyping and development across multiple PLs for multi-platform software~\cite{multiplatform-1, multiplatform-2, multiplatform-3}.
However, achieving accurate automated code translation remains a challenging task due to the inherent differences in syntax, features, and APIs across various PLs~\cite{icse24translationllmsurvey}.

Over the past few decades, various approaches have been proposed to automate code translation and address the associated challenges~\cite{cxcodetranssurvey,ase23ctstudysjtu}. Earlier methods primarily relied on statistical or neural machine translation \cite{fse13statisticalcodetranslation, Tree-to-tree, TransCoder, transcoder-ct, multiplatform-1}, while recent practices start leveraging the powerful large language models (LLMs) as code translators \cite{icse24translationllmsurvey,fse24llmtranslation}.
However, the state-of-the-art practices with LLMs mainly use a direct code-to-code translation approach, where the original program is taken as input to generate the translated program in a single step \cite{fse24llmtranslation}. 
In this manner, LLMs may perform multiple tasks implicitly, such as understanding the semantics of the original program and implementing the semantics into the target program in another PL. Studies have shown that such direct translation remains challenging, notably across PLs with significant differences \cite{icse24translationllmsurvey,icse25intertrans}.

\begin{figure}[t]
\centering
\includegraphics[width=\textwidth]{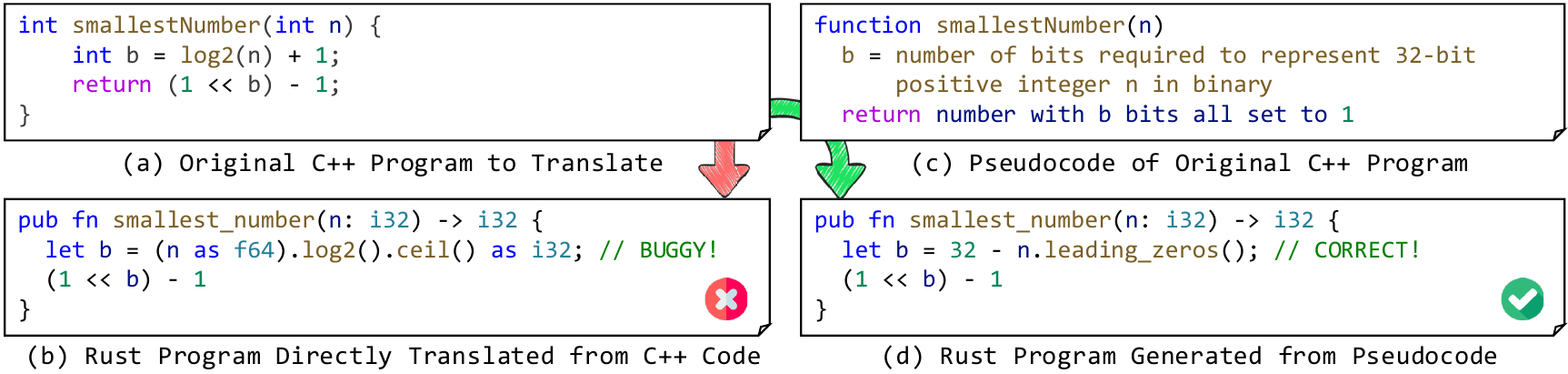}

\setlength{\abovecaptionskip}{5pt}
\caption{An Example of Correct C++-to-Rust Translation by Qwen2.5-Coder-32B-Instruct based on Pseudocode, where Direct Translation Failed. (The other four studied LLMs show similar symptoms in this case.)}
\label{fig: motivation-cpp2rust}
\end{figure}

Decomposing challenging tasks into intermediate steps has proven effective in guiding LLMs to emulate successful human workflows in various coding tasks (e.g., splitting planning and implementation for code generation \cite{mascodegenpku, cotcodegenpku} and fault localization and patch generation for debugging \cite{agentless}). 
However, there are few experiences about whether decomposition also helps generate accurate code translations.
Notably, we observe that \emph{code translation can also benefit from explicitly implementing intermediate steps that emulate human practices in translation}.
Specifically, human translators often perform \emph{semantic translation} \cite{newmark1981semantictranslation} based on the semantics of the given text, first interpreting its meaning and then rendering the meaning in the target language \cite{newmark1981semantictranslation}. 
In addition, a recent study demonstrates that pseudocode, as a PL-agnostic representation of code intent and logic widely used in textbooks and research papers, can effectively guide code generation across PLs \cite{pseudoeval}.
Based on these insights, we realize that emulating semantic translation for code based on pseudocode can be a solution for the tasks that direct translation struggles with. 
For example, when translating a simple C++ program (\Cref{fig: motivation-cpp2rust}(a)) to Rust, all five studied LLMs failed to produce an accurate translation via a one-step direct translation (\Cref{fig: motivation-cpp2rust}(b)). Meanwhile, following a clear pseudocode (\Cref{fig: motivation-cpp2rust}(c)) summarized from the original C++ program, these LLMs can successfully implement the original program's functionality in Rust (\Cref{fig: motivation-cpp2rust}(d)). 
Nevertheless, \textit{the effectiveness, advantages, and limitations of the promising pseudocode-based code translation approach compared to the commonly adopted direct translation remain unclear}.

In this work, we conduct the first systematic empirical study on pseudocode-based code translation to bridge the gap in understanding this under-explored approach. 
Specifically, we assessed the performance of five translation strategies: the widely-adopted direct translation approach, two pseudocode-based strategies, and two hybrid strategies that combine direct and pseudocode-based translation. 
Through four research questions (RQs), we investigate the overall effectiveness of pseudocode-based code translation 
compared with the widely-adopted direct translation (RQ1), its helpfulness for varying PL pairs (RQ2), the advantages of pseudocode as an intermediary compared to a concrete PL, which also proved helpful in code translation \cite{icse25intertrans} (RQ3), and the impact of pseudocode quality on translation performance (RQ4). 
We also examine the successes and failures of pseudocode-based code translation via case studies and discuss future directions for further harnessing its potential benefits.
The study is conducted with solution programs in six popular PLs for 323 LeetCode problems in three difficulty levels in LiveCodeBench \cite{livecodebench}, resulting in 9,690 translation tasks (i.e., 6 source PLs $\times$ 5 target PLs $\times$ 323 problems), for five popular LLMs.

Our experimental results reveal several interesting findings and actionable insights. 
Specifically, we identify the complementary role of pseudocode-based code translation in effectively handling translation tasks that direct translation struggles with. 
By combining the results of direct translation and pseudocode-based translation, the studied LLMs achieve average improvements in translation accuracy over direct translation alone by 4.10\%, 7.44\%, and 13.76\% on easy-, medium-, and hard-level problems, respectively. 
We also demonstrate that pseudocode-based translation can benefit various source-target PL pairs, with more significant benefits for translation from flexible PLs (e.g., Python) to more rigid PLs (e.g., Go), or involving Rust PL that has limited training resources. 
These findings suggest the adoption of a hybrid strategy that combines direct and pseudocode-based translation results with further test-based selection to leverage the complementary strengths of both approaches, particularly when handling the highly benefited PLs/PL pairs.
Moreover, with higher-quality pseudocode, we observed that the studied LLMs can achieve pass rates of 0.9646--0.9835, 0.8861--0.9512, and 0.6747--0.8286 on three-level tasks, respectively, identifying bottlenecks of pseudocode-based code translation in both code understanding and code generation capabilities of LLMs. 
Our case studies further illustrate the advantages of pseudocode-based translation in decoupling the code understanding and generation burdens on complicated programs and mitigating distractions from the PL-specific details in the original programs, as well as its limitations caused by incorrect, incomplete, or ambiguous pseudocode.
These findings illuminate the effective adoption of pseudocode-based translation to enhance LLMs' code translation accuracy and provide evidence for future research on improving LLM-driven code translation.

To summarize, this work makes the following contributions: 
\begin{itemize}[leftmargin=*]
    \item We conduct the first systematic empirical study on pseudocode-based code translation, exploring the effectiveness, advantages, and limitations of emulating human semantic code translation practices via pseudocode in improving code translation accuracy of LLMs. 

    \item We compare four pseudocode-based translation strategies with direct translation across 9,690 translation tasks involving six popular PLs. With five popular LLMs as the translator, we systematically investigate the \textit{effectiveness} of pseudocode-based code translation in enhancing direct translation, its \textit{helpfulness for varying PL pairs}, and its \textit{advantages and limitations}, aiming to provide insights for its effective adoption and evidence for its future enhancement.
    
    \item Our study reveals intriguing findings, including the complementary role of pseudocode-based translation in enhancing direct code translation, the highly benefited PLs/PL pairs, and the potential and bottlenecks of pseudocode-based code translation. These findings yield actionable insights, including the adoption of a hybrid strategy that combines direct and pseudocode-based translation results to enhance code translation accuracy, as well as identify improvement space in the generation of high-quality pseudocode to further harness the benefits of the approach.
\end{itemize}

We release our experimental data, including 1,938 solution programs in six popular PLs collected for 323 LeetCode problems in LiveCodeBench-v5 to translate, as well as scripts to conduct direct and pseudocode-based translation and evaluate multilingual translation results. We will open-source them in our artifact \cite{artifact} to facilitate future research.

The rest of the paper is organized as follows. \Cref{sec:premilinaries} introduces the background of code translation and motivating examples. \Cref{sec:methodology} presents the design of our empirical study, including the research questions, studied code translation strategies, tasks, LLMs, and evaluation metrics. \Cref{sec:evaluation} analyzes the experimental results to answer our research questions in detail and discusses the advantages and limitations of pseudocode-based code translation through case studies. \Cref{sec:threats} and \Cref{sec:relatedwork} discuss the threats to validity and related work, respectively. Finally, \Cref{sec:conclusion} concludes the paper.

\section{Background and Motivation}
\label{sec:premilinaries}

\subsection{Code Translation}
\label{subsec:codetransbg}
Code translation refers to the process of converting programs written in a programming language (PL) into another while preserving their code semantics \cite{fse13statisticalcodetranslation,TransCoder,icse24translationllmsurvey}. 
Formally, given an original program $S = \langle s_1, s_2, \dots, s_n \rangle$ written in the source PL $L_s$, code translation aims to produce a translated program $T = \langle t_1, t_2, \dots, t_m \rangle$ in the target PL $L_t$, such that $T$ performs the same functionality as $S$, where $s_i$ and $t_j$ represent statements in $L_s$ and $L_t$, respectively.
For example, we can translate the C++ program in \Cref{fig: motivation-cpp2rust}(a) and the Python program in \Cref{fig: motivation-py2java}(a) to their semantically equivalent Rust program in \Cref{fig: motivation-cpp2rust}(d) and Java program in \Cref{fig: motivation-py2java}(d), respectively.
In code translation, there may not always be equivalent constructs, statements, and APIs in the target PL for those in the source PL, and vice versa, which makes code translation challenging \cite{fse24llmtranslation,icse25intertrans}.

In the past decade, automated code translation approaches have developed from traditional rule-based and statistical machine translation \cite{fse13statisticalcodetranslation} to neural-model-based methods to automatically learn diverse and complex translation patterns across PLs \cite{Tree-to-tree, TransCoder, transcoder-ct, multiplatform-1, ase23ctstudysjtu}. Recently, LLMs pre-trained on massive code corpora demonstrate superior capabilities in code translation \cite{fse24llmtranslation,fse25alphatrans,icse24translationllmsurvey}. These works mainly focus on a direct one-step code-to-code translation approach, i.e., directly converting the original program in $L_s$ to the translated program in $L_t$ in an integrated step. However, LLMs may struggle to precisely capture and replicate the semantics of the original program in another PL in a single step, leading to incorrect translations \cite{icse24translationllmsurvey}.
To mitigate the gap between source and target PLs, \citet{icse25intertrans} explored transitive translation via an intermediate PL, i.e., chaining two code-to-code translations. But they still fail to handle certain translation tasks. 
In this study, we explore an alternative code translation approach by emulating human semantic translation in natural language translation \cite{newmark1981semantictranslation}, which explicitly conducts code interpretation by generating pseudocode as an intermediate step to understand code intent and logic to facilitate translation.

A property of the code translation task is that the original program is available, enabling automatic correctness assessment of multiple translation candidates. Specifically, unlike the code generation tasks where the result correctness is typically evaluated using manually prepared test assertions or human judgment \cite{humaneval,mbpp}, code translation allows automatic correctness assessment of the translation results by comparing the outputs of the translated and original programs on the same test inputs.
This property makes it practical and meaningful to generate multiple translation candidates to increase the likelihood of getting a correctly translated program, since the original program can help identify the best candidate as the final result via automatic correctness assessment \cite{TransCoder,icse25intertrans,codegeex}.
In this study, we adopt ten generation attempts, which represents a common cost-effective trade-off between computational cost and performance gains on coding tasks \cite{icse25intertrans,pseudoeval}. The multiple-attempt setup also allows us to easily explore the effectiveness of combining results from both direct and pseudocode-based translation approaches (\Cref{subsec: promptingstrategy}).

\subsection{Motivating Examples}
\label{subsec:motivatingexample}

\begin{figure}[t]
\centering

\includegraphics[width=\textwidth]{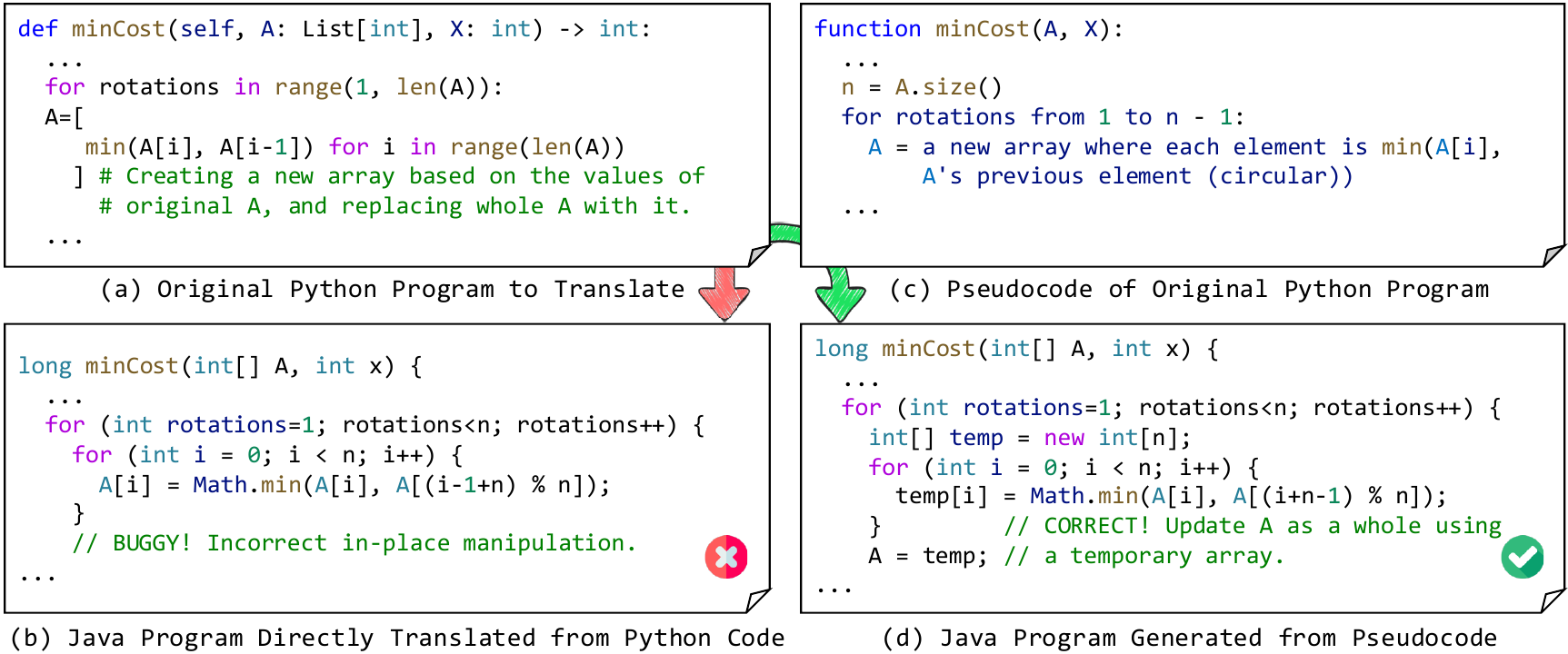}

\setlength{\abovecaptionskip}{5pt}
\caption{An Example of Correct Python-to-Java Translation by DeepSeek-Coder-V2-Lite based on Pseudocode}
\label{fig: motivation-py2java}
\end{figure}

To guide LLMs to solve complicated coding tasks, researchers propose to decompose a complex task into multiple simpler sub-tasks following human experience (e.g., problem solving/planning and implementation for code generation \cite{mascodegenpku, cotcodegenpku, pseudoeval}). Inspired by this, we explore whether decomposing code translation into sub-tasks following human practices can also facilitate LLMs to produce more accurate code translations by emulating the successful human workflow.

Recall the human practices in natural language translation.
When the expressions or sentence structures in source and target languages are sufficiently similar, a literal word-by-word translation is often feasible. 
Meanwhile, when human translators struggle with the significant differences in grammar, idiomatic usage, or context between source and target languages, humans tend to adopt \emph{semantic translation} \cite{newmark1981semantictranslation}, where they first interpret the underlying meaning of the source sentence and then render it in the target language in the manner best to convey the intended message. 

\emph{Interestingly, we observe that \textbf{semantic interpretation also benefits LLMs in code translation}}. 
Specifically, we prompt LLMs to explicitly conduct \textit{semantic translation via pseudocode-based transitive translation}, i.e., first producing a semantic interpretation (pseudocode) for code intent and logic of the original program, and then generating the translated program based on pseudocode. We use pseudocode to represent semantics because it is a common form for describing code intent and logic of algorithms in textbooks and research papers. Also, it proves effective in guiding LLMs to replicate the code intent and logic in varying PLs and can be synthesized from programs~\cite{pseudoeval}.

{\Cref{fig: motivation-py2java} shows an example where pseudocode enables DeepSeek-Coder-V2-Lite {(\textit{abbr.} DSCoder)} to correctly translate a Python program that cannot be accurately transformed through direct translation. Specifically, when translating the Python program in \Cref{fig: motivation-py2java}(a) into Java {directly}, DSCoder attempts to mimic the Python implementation by using a for loop. Although it recognizes that Java does not support lambda expressions within for loops, the resulting translation (\Cref{fig: motivation-py2java}(b)) employs in-place manipulation that is structurally similar but semantically inconsistent with the original Python program. 
In contrast, when DSCoder first abstracts the intent of the Python program into pseudocode (\Cref{fig: motivation-py2java}(c)) and then generates Java code from this pseudocode, it produces a correct implementation by updating the array as a whole via a temporary array (\Cref{fig: motivation-py2java}(d)).}

Moreover, we observe that the semantic code translation via pseudocode is broadly helpful for LLMs in various families and sizes. 
As shown in \Cref{fig: motivation-cpp2rust}, when translating a simple C++ solution of an easy-level LeetCode programming problem into Rust, the powerful Qwen2.5-Coder-32B-Instruct (\textit{abbr.} Qwen32B) fails to produce a correct translation via direct translation, by misusing the \texttt{log2} API in Rust as shown in \Cref{fig: motivation-cpp2rust}(b). However, when guided by the corresponding pseudocode of the original program (\Cref{fig: motivation-cpp2rust}(c)), Qwen32B correctly leverages the \texttt{leading\_zeros} API to achieve the intended functionality as shown in \Cref{fig: motivation-cpp2rust}(d). Notably, all five studied LLMs (listed in \Cref{table: llminfo}) suffer from similar issues, and all of them except DSCoder generate a correct implementation when provided with pseudocode summarized by themselves as an intermediate step. 

These motivating examples suggest that \textit{{incorporating pseudocode as a semantic representation and conducting transitive translation can promote accurate code translations}}, mirroring the benefits of humans using semantic translation for natural languages. 
This is reasonable as the approach guides LLMs to understand code intent and logic and then perform code generation, decomposing a complex task into simpler sub-tasks, and LLMs are skillful in both tasks \cite{llmgoodsummarizer,livecodebench}. This enables a new access to achieving code translation, which may bypass LLMs' struggle with direct translation.

Despite these benefits, the proper use of pseudocode in code translation remains underexplored, e.g., how to effectively translate code via pseudocode, how effective it is when handling different source-target PL pairs, and what strengths it entails.
In addition, pseudocode-based code translation may also face challenges, e.g., the potential error propagation from incorrect pseudocode and the inherent ambiguity in natural language \cite{incorrectcot}. 
The limitations it entails and potential remedies to overcome the limitations also remain unknown. To bridge these gaps, we conduct a systematic investigation of pseudocode-guided code translation. Our study aims to provide insights into how emulating semantic translation via pseudocode can effectively enhance code translation. 

\section{Study Design}
\label{sec:methodology}

\subsection{Research Questions}

In this study, we investigate the effectiveness of pseudocode-based code translation by studying the following four research questions (RQs).

\textbullet{ } \textbf{RQ1: How effective is pseudocode-based code translation?}

This RQ aims to show a general picture of the effectiveness of emulating semantic translation for code via pseudocode. We investigate {whether pseudocode-based translation effectively complements direct translation and whether it can fully replace direct translation, aiming to shed light on the effective strategy to leverage translation approaches.}

\textbullet{ } \textbf{RQ2: How helpful is pseudocode-based code translation for different pairs of source and target programming languages?} 

Existing studies show that the code translation accuracy of LLMs varies across different pairs of source and target PLs, indicating that distinct PL pairs may pose varying challenges for LLMs \cite{icse24translationllmsurvey,icse25intertrans}. Pseudocode may also have varying helpfulness for different source and target PL pairs.
Therefore, this RQ verifies whether pseudocode-based translation is generally helpful for all PL pairs. It also investigates whether certain PL pairs can benefit more from pseudocode-based translation, suggesting scenarios where pseudocode-based translation is highly recommended.

\textbullet{ } \textbf{RQ3: How effective is pseudocode as an intermediary compared to a specific programming language?} 

\citet{icse25intertrans} demonstrates that a specific PL with close syntactic and semantic similarities to the source and target PLs also benefits transitive code translation. For example, translating Python code to Java via a Rust implementation as a bridge is found to complement direct Python-to-Java translation. In this RQ, we aim to understand whether pseudocode, which interprets mainly code intent and logic without PL-specific details, can serve as a more effective intermediary than a specific PL in transitive code translation. 

\textbullet{ } \textbf{RQ4: How does the quality of pseudocode affect the effectiveness of pseudocode-based code translation?}

LLMs may not accurately interpret the intent and logic of original programs, which may result in low-quality pseudocode, conversely hindering LLMs from generating correctly translated programs. 
This RQ investigates the code translation performance based on high-quality pseudocode, aiming to reveal the potential of pseudocode-based code translation and identify the bottlenecks of studied LLMs to harness the potential effectiveness of pseudocode-based code translation.

\subsection{Experimental Translation Subjects}
\label{subsec:plandtranstasks}

\subsubsection{Programming Languages}
{
In this study, we evaluate code translation performance across six widely studied PLs, i.e., C++, Python, Java, JavaScript, Go, and Rust. The choice of these PLs follows prior work, where translation among the first four popular PLs \cite{tiobeindex} has been extensively explored~\cite{transcoder-ct, ase23ctstudysjtu, fse24llmtranslation}, while recent studies have highlighted the emerging Go and Rust~\cite{icse24translationllmsurvey, icse25intertrans}.
}

\subsubsection{Code Translation Tasks}
We prompt LLMs to translate the solution programs of 323 LeetCode problems in LiveCodeBench{-v5} \cite{livecodebench}. 
We choose these programs because they involve common implementations widely used in practical development and form a representative benchmark to evaluate the coding ability of LLMs \cite{qwen25techreport,dpsv3techreport}. 
Actually, solution programs of LeetCode problems have been widely adopted as subjects in code translation studies, as they provide easy-to-collect multilingual parallel programs \cite{oopslatranspilation,avataracl}.
Besides, LiveCodeBench offers carefully validated and representative tests, which facilitate the validation of translation results. Additionally, the LiveCodeBench programming problems are released after May 2023 \cite{livecodebench}, allowing the evaluation to be less subject to the data contamination issue compared to the conventional code translation benchmarks like CodeNet \cite{CodeNet}, TransCoder \cite{TransCoder}, and AVATAR \cite{avataracl} established in 2020--2021. 

\begin{table}[t]
\setlength{\tabcolsep}{4pt}
\caption{AST Node Counts and API Invocation Counts of Programs for Different Difficulty-Level Problems\label{table: programdiffinfo}}
\footnotesize
\begin{tabular}{c|cccccc|cccccc}
\toprule
\multirow{2}{*}{\textbf{Difficulty}} & \multicolumn{6}{c|}{\textbf{Average Count of AST   Nodes}}                                                                                                                                                                                                                                                                                                & \multicolumn{6}{c}{\textbf{Average Count of   Invocation Nodes}}                                                                                                                                                                                                                                                                                 \\
                                     & \textbf{C++}                                            & \textbf{Go}                                             & \textbf{Java}                                           & \textbf{JS}                                             & \textbf{Python}                                         & \textbf{Rust}                                           & \textbf{C++}                                           & \textbf{Go}                                           & \textbf{Java}                                         & \textbf{JS}                                            & \textbf{Python}                                       & \textbf{Rust}                                          \\ \midrule
Easy                                 & \begin{tabular}[c]{@{}c@{}}167.91\\ \textit{(1.0x)}\end{tabular} & \begin{tabular}[c]{@{}c@{}}142.81\\ \textit{(1.0x)}\end{tabular} & \begin{tabular}[c]{@{}c@{}}159.33\\ \textit{(1.0x)}\end{tabular} & \begin{tabular}[c]{@{}c@{}}138.50\\ \textit{(1.0x)}\end{tabular} & \begin{tabular}[c]{@{}c@{}}111.22\\ \textit{(1.0x)}\end{tabular} & \begin{tabular}[c]{@{}c@{}}173.58\\ \textit{(1.0x)}\end{tabular} & \begin{tabular}[c]{@{}c@{}}2.56\\ \textit{(1.0x)}\end{tabular}  & \begin{tabular}[c]{@{}c@{}}2.29\\ \textit{(1.0x)}\end{tabular} & \begin{tabular}[c]{@{}c@{}}2.94\\ \textit{(1.0x)}\end{tabular} & \begin{tabular}[c]{@{}c@{}}1.97\\ \textit{(1.0x)}\end{tabular}  & \begin{tabular}[c]{@{}c@{}}3.32\\ \textit{(1.0x)}\end{tabular} & \begin{tabular}[c]{@{}c@{}}5.16\\ \textit{(1.0x)}\end{tabular}  \\
Medium                               & \begin{tabular}[c]{@{}c@{}}280.25\\ \textit{(1.7x)}\end{tabular} & \begin{tabular}[c]{@{}c@{}}263.34\\ \textit{(1.8x)}\end{tabular} & \begin{tabular}[c]{@{}c@{}}267.83\\ \textit{(1.7x)}\end{tabular} & \begin{tabular}[c]{@{}c@{}}266.81\\ \textit{(1.9x)}\end{tabular} & \begin{tabular}[c]{@{}c@{}}188.18\\ \textit{(1.7x)}\end{tabular} & \begin{tabular}[c]{@{}c@{}}298.54\\ \textit{(1.7x)}\end{tabular} & \begin{tabular}[c]{@{}c@{}}5.43\\ \textit{(2.1x)}\end{tabular}  & \begin{tabular}[c]{@{}c@{}}4.95\\ \textit{(2.2x)}\end{tabular} & \begin{tabular}[c]{@{}c@{}}5.46\\ \textit{(1.9x)}\end{tabular} & \begin{tabular}[c]{@{}c@{}}4.42\\ \textit{(2.2x)}\end{tabular}  & \begin{tabular}[c]{@{}c@{}}5.64\\ \textit{(1.7x)}\end{tabular} & \begin{tabular}[c]{@{}c@{}}8.69\\ \textit{(1.7x)}\end{tabular}  \\
Hard                                 & \begin{tabular}[c]{@{}c@{}}592.81\\ \textit{(3.5x)}\end{tabular} & \begin{tabular}[c]{@{}c@{}}529.85\\ \textit{(3.7x)}\end{tabular} & \begin{tabular}[c]{@{}c@{}}519.15\\ \textit{(3.3x)}\end{tabular} & \begin{tabular}[c]{@{}c@{}}546.64\\ \textit{(3.9x)}\end{tabular} & \begin{tabular}[c]{@{}c@{}}339.70\\ \textit{(3.1x)}\end{tabular} & \begin{tabular}[c]{@{}c@{}}584.04\\ \textit{(3.4x)}\end{tabular} & \begin{tabular}[c]{@{}c@{}}10.55\\ \textit{(4.1x)}\end{tabular} & \begin{tabular}[c]{@{}c@{}}8.85\\ \textit{(3.9x)}\end{tabular} & \begin{tabular}[c]{@{}c@{}}7.43\\ \textit{(2.5x)}\end{tabular} & \begin{tabular}[c]{@{}c@{}}11.23\\ \textit{(5.7x)}\end{tabular} & \begin{tabular}[c]{@{}c@{}}8.77\\ \textit{(2.6x)}\end{tabular} & \begin{tabular}[c]{@{}c@{}}13.68\\ \textit{(2.6x)}\end{tabular} \\ \bottomrule
\end{tabular}
\end{table}

To support the investigation of code translation among diverse PLs, we follow existing practices \cite{livecodebench,pseudoeval} to collect user-written solution programs that can pass all the tests on LiveCodeBench, for all six PLs. 
These solution programs are used as the original programs to translate.
Among all 381 LeetCode problems from LiveCodeBench-v5, we excluded 16 problems with incorrect tests identified by \citet{pseudoeval}. 
We also excluded another 42 problems lacking solutions in at least one of the six PLs on LeetCode. 
This results in 116 easy problems, 160 medium problems, and 47 hard problems. 
{The solution programs of harder problems typically include more complex solving logic and implementations \cite{livecodebench}. 
As shown in \Cref{table: programdiffinfo}, the average abstract syntax tree (AST) node count and API invocation count\footnote{The statistics are based on counts of all AST nodes and external function calls identified by \textit{tree-sitter}~\cite{tree-sitter}, respectively.}
of solution programs for medium/hard-level problems are 1.7x--1.9x / 3.1x--3.9x and 1.7x--2.2x / 2.5x--5.7x as those of easy-level problems across six PLs, suggesting that the harder tasks present higher translation complexity \cite{repoleveltransarxivsyt}.}
Note that, for each problem, LiveCodeBench does not provide corresponding solution programs and only supports the evaluation of Python solutions.
We collect the solution programs in six PLs and extend the execution-based evaluation scaffold to all six studied PLs.

Finally, we constructed a code translation dataset based on 323 LeetCode problems in LiveCodeBench. 
For each problem, there are six validated solution programs in distinct PLs and corresponding tests. 
Each validated solution program is taken in turn as the original program for translation into the other five PLs, formulating 9,690 translation tasks involving 6$\times$5=30 source-target PL pairs. 
All LLM-translated programs in target PLs will be evaluated by tests provided by LiveCodeBench \cite{livecodebench}.

\subsection{Experimental Strategies for Code Translation}
\label{subsec: promptingstrategy}

\begin{figure}[t]
\centering
\begin{subfigure}{0.4925\textwidth}
\includegraphics[width=\textwidth]{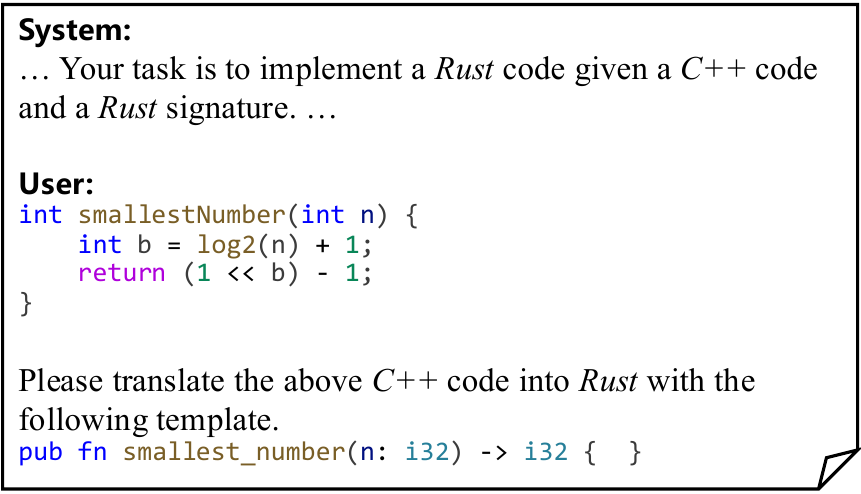}
\subcaption[]{Prompt for Direct Translation\label{subfig: prompt-directtranslate}}
\end{subfigure}
\,
\begin{subfigure}{0.4915\textwidth}
\includegraphics[width=\textwidth]{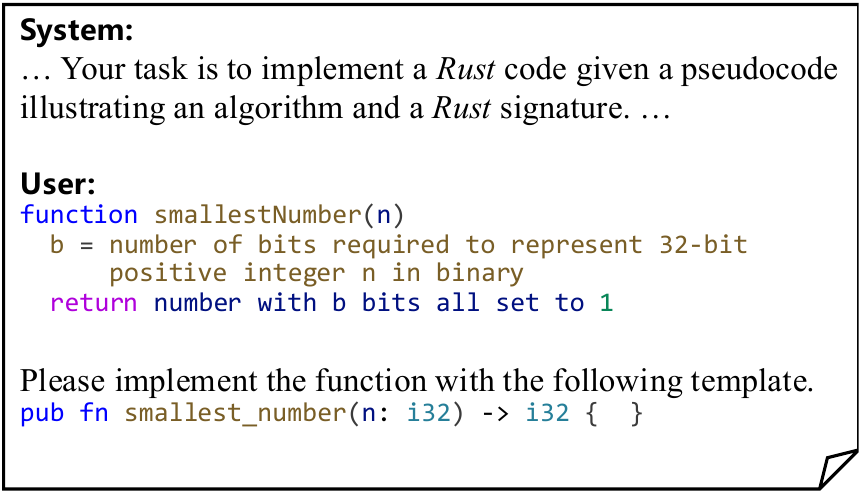}
\subcaption[]{Prompt for Generating Code from Pseudocode\label{subfig: prompt-pseudocodewocode}}
\end{subfigure}

\begin{subfigure}{\textwidth}
\includegraphics[width=\textwidth]{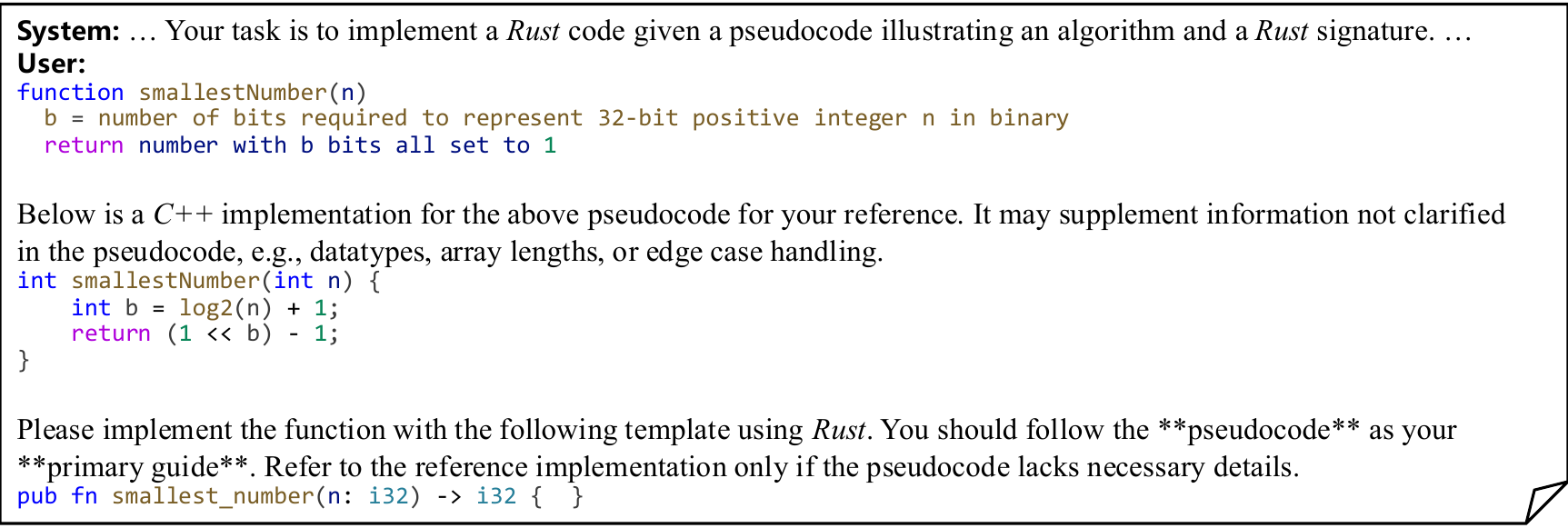}
\subcaption[]{Prompt for Generating Code from Pseudocode with Original Program as Reference Implementation\label{subfig: prompt-pseudocodewcode}}
\end{subfigure}

\caption{Prompts Used in the Study (The examples are instantiated with the translation tasks in \Cref{fig: motivation-cpp2rust}).}
\end{figure}

In this study, we investigate the effectiveness of five prompting strategies for code translation. These include one strategy based on the commonly adopted direct translation approach, two pseudocode-based translation strategies, and two hybrid translation strategies.

\begin{enumerate}[leftmargin=*]
    \item \textit{Direct Translation (abbr. \sd):} This strategy prompts LLMs to translate original programs in the source PL to the target PL in one step. The strategy is widely used by existing LLM-based code translation studies \cite{fse24llmtranslation,icse24translationllmsurvey}. We follow these studies to design the prompt as shown in \Cref{subfig: prompt-directtranslate}.

    \item \textit{Transitive Translation via Pseudocode (abbr. \svp):} This strategy prompts LLMs to emulate semantic translation using two transitive steps explicitly: (i) generate a piece of pseudocode to describe the PL-agnostic code intent and logic for the original program; and (ii) implement the pseudocode into a program in the target PL following the prompt in \Cref{subfig: prompt-pseudocodewocode}. 
    As introduced in \Cref{subsec:motivatingexample}, this strategy is inspired by human practices of translation and the findings that pseudocode can encode the PL-agnostic code intent and logic of a program \cite{pseudoeval}, serving as an appropriate semantic representation for emulating semantic translation. 
    In our study, pseudocode and translated programs are generated by the same LLM (referred to as ``\emph{translator LLM}''). Our prompts are designed following practices of \citet{pseudoeval}. 
    In particular, we reuse their prompt for pseudocode generation, which guides LLMs to generate pseudocode following: 1) using indents to represent control structures; 2) avoiding usages and type information specific to PLs; 3) omitting implementation of common algorithms, data structures, and operations; and 4) naming functions using the template \code{function [FUNC\_NAME] ([PARAM\_LIST])} (e.g., \code{function max(a, b)}). These conventions align with the general expectations of pseudocode style in prior studies \cite{Kul19,iclr25pcoderetrieval,Ach22}. Given that there is no unified format and grammar for pseudocode, we allow LLMs to generate pseudocode in their most familiar style based on these conventions. This prompt is available in our artifact \cite{artifact}.

    \item \textit{Transitive Translation via Pseudocode with Original Program as Context (abbr. \svpp):} 
    This strategy extends the above pseudocode-based strategy to include implementation details (i.e., original program) as the context to guide code translation. 
    Specifically, considering that pseudocode abstracts implementation details of the original program, it may miss some helpful information for LLMs to implement the original program's intent and logic in the target PL. Thus, we include the original program in the source PL as an implementation example of the pseudocode for LLMs' reference when generating the translated program by emulating semantic translation, using the prompt shown in \Cref{subfig: prompt-pseudocodewcode}.
\end{enumerate}
As mentioned in \Cref{subsec:codetransbg}, we generate ten translations for each original program. 
When using the above strategies, we repeatedly generate ten translations using their respective prompts.

We also examine the effectiveness of combining direct translation and pseudocode-based translation. 
Specifically, we design two straightforward hybrid strategies by evenly mixing direct translation results and pseudocode-based translation results. We keep the same total attempts of ten as the above three single-approach strategies, thus combining five translated programs generated by each approach.
The two hybrid strategies are defined as follows:
\begin{enumerate}[leftmargin=*]
    \item[(4)] \textit{Hybrid translation \sdvp{}:} This strategy mixes five translation results from \textit{Direct Translation} (\sd) and another five translated results from \textit{Transitive Translation via Pseudocode} (\svp). 
    \item[(5)] \textit{Hybrid translation \sdvpp{}:} This strategy mixes five translation results from \textit{Direct Translation} (\sd) and another five translated results from \textit{Transitive Translation via Pseudocode with Original Program as Context} (\svpp). 
\end{enumerate}
We include these hybrid strategies to study whether combining direct translation and pseudocode-based translation can leverage their complementary strengths across different translation tasks.

\subsection{Experimental Setups}

\noindent \textbf{Metric:} We report the \textit{pass@10 rate} of the translated programs (i.e., programs in the target PL) on the test cases provided by LiveCodeBench to reflect the computational accuracy of the translation results. The metric measures the ratio of tasks on which the translator LLM successfully generates a correct translation in at least one out of ten attempts. 
Compared with text-based metrics like BLEU and Exact Match rate, the execution-based \textit{pass@10} metric can measure the semantic equivalence of the translation results based on the tests provided by LiveCodeBench. 
The metric is widely adopted in code generation \cite{livecodebench,pseudoeval} and translation studies \cite{icse25intertrans}. 
Note that we use \textit{pass@10} to investigate the potential of the studied translation strategies and LLMs, considering the higher translation accuracy achieved with ten attempts than one single attempt demonstrated by existing study \cite{icse25intertrans} and the practical feasibility of filtering translation results by referring to the original program's behavior as we discuss in \Cref{subsec:codetransbg}. In addition, we discuss \textit{pass@\{2,4,6,8\}} in RQ1.

\noindent \textbf{Studied LLMs:}
In this study, we investigate five popular LLMs deployable on our machine.
They are {Qwen2.5-Coder-32B-Instruct \textit{(abbr. Qwen32B)}} \cite{qwen25techreport}, {Qwen2.5-Coder-7B-Instruct \textit{(abbr. Qwen7B)}} \cite{qwen25techreport}, {Phi-4} \cite{phi4techreport}, {GPT-4o-mini} \cite{urlgpt4omini}, and {DeepSeek-Coder-V2-Lite-Instruct \textit{(abbr. DSCoder)}} \cite{dpsv2coderpaper}. 
They include both general-purpose LLMs and coding LLMs, and both open-sourced and commercial LLMs from different families. The detailed information of these LLMs is listed in \Cref{table: llminfo}.

\begin{table}[t]
\caption{Information of LLMs Involved in the Study\label{table: llminfo}}
\footnotesize
\begin{tabular}{ccccc}
\toprule
\textbf{LLM Name}                                                                         & \textbf{Family} & \textbf{Type} & \textbf{Parameter Size} & \textbf{Knowledge Cutoff} \\ \midrule
\begin{tabular}[c]{@{}c@{}}Qwen2.5-Coder-32B-Instruct*\\ \textit{(abbr. Qwen32B)}\end{tabular}      & Qwen                  & Open Source   & 32.8B                   & Mar 2024 \cite{qwen32b-cufoff}                 \\ \midrule
Phi-4                                                                                     & Microsoft             & Open Source   & 14.7B                   & May 2024 \cite{phi4-cufoff}                 \\ \midrule
GPT-4o-mini                                                                               & OpenAI                & Commercial    & (Unknown)               & Oct 2023 \cite{gpt4omini-cufoff}                 \\ \midrule
\begin{tabular}[c]{@{}c@{}}Qwen2.5-Coder-7B-Instruct\\ \textit{(abbr. Qwen7B)}\end{tabular}        & Qwen                  & Open Source   & 7.62B                   & Mar 2024 \cite{qwen7b-cufoff}                 \\ \midrule
\begin{tabular}[c]{@{}c@{}}DeepSeek-Coder-V2-Lite\\ -Instruct \textit{(abbr. DSCoder)}\end{tabular} & DeepSeek              & Open Source   & 15.7B                   & Nov 2023 \cite{dscoder-cufoff}                \\ \bottomrule
\multicolumn{5}{p{13cm}}{{\scriptsize *: We run Qwen2.5-Coder-32B-Instruct-GPTQ-Int4 with NVIDIA RTX4090 GPU. The quantized GPTQ-Int4 version is found to show comparable coding ability as the original bf16 version while taking less GPU memory \cite{pseudoeval}.}}

\end{tabular}
\end{table}

\noindent \textbf{LLM Configuration:} We run the four open-source LLMs on our machine and access the commercial GPT-4o-mini via OpenAI API. 
To balance the diversity and reliability, we adopt the temperature of 0.2 following the practice in LLM-based coding studies \cite{livecodebench,pseudoeval, mradopt}. The maximum output tokens are set to be 3000 (representing 1.5x maximum input code length) to accommodate normal translation outputs. The other configurations (e.g., top\_p  and penalty) are kept as default values.
To mitigate randomness, we repeat the experiments three times and report the average results.

\noindent \textbf{Experimental Environment:} We run experiments on a machine with Ubuntu 22.04 OS. The server is equipped with NVIDIA RTX4090 GPUs to deploy the open-source LLMs, as well as an AMD Ryzen Threadripper PRO 3995WX 64-Core CPU to support the parallel evaluation of plenty of generated translation results.

\section{Results and Analysis}
\label{sec:evaluation}

\subsection{RQ1: Overall Effectiveness of Pseudocode-based Code Translation}

In RQ1, we investigate the overall effectiveness of pseudocode-based code translation. 
Specifically, we explore (1) {whether pseudocode-based code translation can advance the accuracy of the current widely-adopted direct translation approach, by either complementarily enhancing or completely replacing direct translation;} 
and (2) whether pseudocode generated by the translator LLM can effectively represent the code intent or logic of the original program under translation and promote accurate code translation results.
We investigate these by comparing the performance of five translation strategies introduced in \Cref{subsec: promptingstrategy}. We take the widely adopted direct translation strategy ({\sd}) as the baseline and analyze the performance improvement of the other four strategies. \cref{table: RQ1-viaPseudocode-withSrc} shows the pass@10 rate of each LLM using different strategies.

\begin{table}[t]
    \setlength{\tabcolsep}{4pt}
    \caption{Pass@10 Rate of Code Translation Using Different Translation
    Strategies\label{table: RQ1-viaPseudocode-withSrc}}
    \footnotesize
    \begin{tabular}{clc@{\hskip 1pt}cc@{\hskip 1pt}cc@{\hskip 1pt}c}
        \toprule \textbf{LLM}                        & \multicolumn{1}{c}{\textbf{Strategy}}                          & \multicolumn{2}{c}{\textbf{Easy Tasks}} & \multicolumn{2}{c}{\textbf{Medium Tasks}} & \multicolumn{2}{c}{\textbf{Hard Tasks}} \\
        \hline
        \multirow{5}{*}{Qwen32B}                     & {\sd}\;\;{\scriptsize (Direct Translation)}                    & 0.9480                                  &                                           & 0.8772                                 &                               & 0.7059          &                                \\
                                                     & {\svp}\;\;{\scriptsize (viaPseudocode)}                        & 0.9442                                  & {\scriptsize(-0.40\%)}                    & 0.8736                                 & {\scriptsize(-0.41\%)}        & 0.6768          & {\scriptsize(-4.12\%)}         \\
                                                     & {\svpp}\;\;{\scriptsize (viaPseudocode w/ original program as context)} & 0.9648                                  & {\scriptsize(+1.77\%)}                    & 0.9001                                 & {\scriptsize(+2.61\%)}        & 0.7553          & {\scriptsize(+7.00\%)}         \\
                                                     & {\sdvp}\;\;{\scriptsize (Hybrid [DT \& viaPseudocode])}            & \textbf{0.9773}                         & \textbf{\scriptsize(+3.09\%)}             & \textbf{0.9403}                        & \textbf{\scriptsize(+7.19\%)} & 0.7948          & {\scriptsize(+12.59\%)}        \\
                                                     & {\sdvpp}\;\;{\scriptsize (Hybrid [DT \& viaPseudocode w/ context])}  & \textbf{0.9773}                         & \textbf{\scriptsize(+3.09\%)}             & 0.9327                                 & {\scriptsize(+6.33\%)}        & \textbf{0.7988} & \textbf{\scriptsize(+13.16\%)} \\
        \midrule \multirow{5}{*}{Phi-4}              & {\sd}\;\;{\scriptsize (Direct Translation)}                    & 0.9276                                  &                                           & 0.8313                                 &                               & 0.6281          &                                \\
                                                     & {\svp}\;\;{\scriptsize (viaPseudocode)}                        & 0.8782                                  & {\scriptsize(-5.33\%)}                    & 0.7744                                 & {\scriptsize(-6.84\%)}        & 0.5832          & {\scriptsize(-7.15\%)}         \\
                                                     & {\svpp}\;\;{\scriptsize (viaPseudocode w/ original program as context)} & 0.9254                                  & {\scriptsize(-0.24\%)}                    & 0.8347                                 & {\scriptsize(+0.41\%)}        & 0.6463          & {\scriptsize(+2.90\%)}         \\
                                                     & {\sdvp}\;\;{\scriptsize (Hybrid [DT \& viaPseudocode])}            & \textbf{0.9646}                         & \textbf{\scriptsize(+3.99\%)}             & \textbf{0.8947}                        & \textbf{\scriptsize(+7.63\%)} & \textbf{0.7083} & \textbf{\scriptsize(+12.77\%)} \\
                                                     & {\sdvpp}\;\;{\scriptsize (Hybrid [DT \& viaPseudocode w/ context])}  & 0.9610                                  & {\scriptsize(+3.60\%)}                    & 0.8903                                 & {\scriptsize(+7.10\%)}        & 0.7040          & {\scriptsize(+12.08\%)}        \\
        \midrule \multirow{5}{*}{\begin{tabular}[c]{@{}c@{}}GPT-4o\\ -mini\end{tabular}}        & {\sd}\;\;{\scriptsize (Direct Translation)}                    & 0.9279                                  &                                           & 0.8552                                 &                               & 0.6667          &                                \\
                                                     & {\svp}\;\;{\scriptsize (viaPseudocode)}                        & 0.8647                                  & {\scriptsize(-6.81\%)}                    & 0.7219                                 & {\scriptsize(-15.59\%)}       & 0.5000          & {\scriptsize(-25.00\%)}        \\
                                                     & {\svpp}\;\;{\scriptsize (viaPseudocode w/ original program as context)} & 0.9239                                  & {\scriptsize(-0.43\%)}                    & 0.8635                                 & {\scriptsize(+0.97\%)}        & 0.6837          & {\scriptsize(+2.55\%)}         \\
                                                     & {\sdvp}\;\;{\scriptsize (Hybrid [DT \& viaPseudocode])}            & \textbf{0.9658}                         & \textbf{\scriptsize(+4.08\%)}             & 0.9054                                 & {\scriptsize(+5.87\%)}        & 0.7404          & {\scriptsize(+11.05\%)}        \\
                                                     & {\sdvpp}\;\;{\scriptsize (Hybrid [DT \& viaPseudocode w/ context])}  & 0.9621                                  & {\scriptsize(+3.69\%)}                    & \textbf{0.9144}                        & \textbf{\scriptsize(+6.92\%)} & \textbf{0.7617} & \textbf{\scriptsize(+14.25\%)} \\
        \midrule \multirow{5}{*}{Qwen7B}             & {\sd}\;\;{\scriptsize (Direct Translation)}                    & 0.9073                                  &                                           & 0.7928                                 &                               & 0.5400          &                                \\
                                                     & {\svp}\;\;{\scriptsize (viaPseudocode)}                        & 0.7735                                  & {\scriptsize(-14.75\%)}                   & 0.5928                                 & {\scriptsize(-25.23\%)}       & 0.3286          & {\scriptsize(-39.15\%)}        \\
                                                     & {\svpp}\;\;{\scriptsize (viaPseudocode w/ original program as context)} & 0.8989                                  & {\scriptsize(-0.93\%)}                    & 0.7929                                 & {\scriptsize(+0.01\%)}        & 0.5480          & {\scriptsize(+1.48\%)}         \\
                                                     & {\sdvp}\;\;{\scriptsize (Hybrid [DT \& viaPseudocode])}            & 0.9336                                  & {\scriptsize(+2.90\%)}                    & 0.8395                                 & {\scriptsize(+5.89\%)}        & 0.5849          & {\scriptsize(+8.31\%)}         \\
                                                     & {\sdvpp}\;\;{\scriptsize (Hybrid [DT \& viaPseudocode w/ context])}  & \textbf{0.9420}                         & \textbf{\scriptsize(+3.82\%)}             & \textbf{0.8568}                        & \textbf{\scriptsize(+8.07\%)} & \textbf{0.6154} & \textbf{\scriptsize(+13.96\%)} \\
        \midrule \multirow{5}{*}{DSCoder}            & {\sd}\;\;{\scriptsize (Direct Translation)}                    & 0.8964                                  &                                           & 0.7876                                 &                               & 0.5586          &                                \\
                                                     & {\svp}\;\;{\scriptsize (viaPseudocode)}                        & 0.8562                                  & {\scriptsize(-4.48\%)}                    & 0.7335                                 & {\scriptsize(-6.87\%)}        & 0.5069          & {\scriptsize(-9.26\%)}         \\
                                                     & {\svpp}\;\;{\scriptsize (viaPseudocode w/ original program as context)} & 0.9218                                  & {\scriptsize(+2.83\%)}                    & 0.7875                                 & {\scriptsize(-0.01\%)}        & 0.5530          & {\scriptsize(-1.00\%)}         \\
                                                     & {\sdvp}\;\;{\scriptsize (Hybrid [DT \& viaPseudocode])}            & 0.9482                                  & {\scriptsize(+5.78\%)}                    & 0.8569                                 & {\scriptsize(+8.80\%)}        & 0.6442          & {\scriptsize(+15.32\%)}        \\
                                                     & {\sdvpp}\;\;{\scriptsize (Hybrid [DT \& viaPseudocode w/ context])}  & \textbf{0.9537}                         & \textbf{\scriptsize(+6.39\%)}             & \textbf{0.8581}                        & \textbf{\scriptsize(+8.95\%)} & \textbf{0.6459} & \textbf{\scriptsize(+15.63\%)} \\
        \midrule \multirow{5}{*}{\textit{(Average)}} & {\sd}\;\;{\scriptsize (Direct Translation)}                    & 0.9214                                  &                                           & 0.8288                                 &                               & 0.6199          &                                \\
                                                     & {\svp}\;\;{\scriptsize (viaPseudocode)}                        & 0.8634                                  & {\scriptsize(-6.30\%)}                    & 0.7392                                 & {\scriptsize(-10.81\%)}       & 0.5191          & {\scriptsize(-16.26\%)}        \\
                                                     & {\svpp}\;\;{\scriptsize (viaPseudocode w/ original program as context)} & 0.9270                                  & {\scriptsize(+0.60\%)}                    & 0.8357                                 & {\scriptsize(+0.83\%)}        & 0.6373          & {\scriptsize(+2.81\%)}         \\
                                                     & {\sdvp}\;\;{\scriptsize (Hybrid [DT \& viaPseudocode])}            & 0.9579                                  & {\scriptsize(+3.96\%)}                    & 0.8874                                 & {\scriptsize(+7.06\%)}        & 0.6945          & {\scriptsize(+12.04\%)}        \\
                                                     & {\sdvpp}\;\;{\scriptsize (Hybrid [DT \& viaPseudocode w/ context])}  & \textbf{0.9592}                         & \textbf{\scriptsize(+4.10\%)}             & \textbf{0.8905}                        & \textbf{\scriptsize(+7.44\%)} & \textbf{0.7052} & \textbf{\scriptsize(+13.76\%)} \\
        \bottomrule                                   
    \end{tabular}
\end{table}

\textit{\textbf{Effectiveness in Enhancing Direct Translation.}}
The results show that the two hybrid strategies (i.e., {\sdvp} and {\sdvpp}) that combine translation outcomes from pseudocode-based and direct translation effectively improved the code translation accuracy achieved by the baseline ({\sd}). 
Specifically, {\sdvp} improved the pass@10 rate by 3.96\%, 7.06\%, and 12.04\% on average on three difficulty levels, respectively. {\sdvpp} led to more significant improvements of 4.10\%, 7.44\%, and 13.76\% on three levels on average, respectively. 
The results demonstrate that pseudocode-based code translation enables LLMs to correctly translate many programs that cannot be correctly handled by direct translation alone. Thus, given the same number of attempts, it is beneficial to adopt both direct translation and pseudocode-based translation instead of consistently giving chances to either strategy. The results indicate that pseudocode-based code translation can effectively complement the widely adopted direct translation to enhance code translation accuracy.

Meanwhile, we observed that the two pure pseudocode-based translation strategies (i.e., {\svp} and {\svpp}) produced only comparable or even worse performance than direct translation ({\sd}). Specifically, {\svpp} increased pass@10 of {\sd} by only 0.60\%, 0.83\%, and 2.81\% on three levels on average; {\svp} even led to drops in pass@10 rate by 6.30\%, 10.81\%, and 16.26\%. The results demonstrate that when conducting semantic translation via pseudocode, LLMs also failed to translate a few programs that could be correctly translated by direct translation. This indicates that direct translation and pseudocode-based translation are complementary at the current stage.
We discuss how this underperformance can be attributed to incorrect, incomplete, and ambiguous pseudocode via case studies in \Cref{subsec: case-study}, and we also explore the potential of pure pseudocode-based code translation based on higher-quality pseudocode in RQ4 (\Cref{subsec:rq4}).

\textit{Difficulty-wise.} We also notice that the improvement of hybrid strategies is more significant on harder tasks. Specifically, the improvement in pass@10 rate brought by {\sdvp} and {\sdvpp} over {\sd} increases by 3.96\% and 4.10\% on easy-level tasks, while 12.04\% and 13.76\% on hard-level tasks on average, respectively. 
{As introduced in \Cref{subsec:plandtranstasks}, the original programs of harder LeetCode problems typically include more complex solving logic and implementations, involving more AST nodes and more external API invocations to translate.}
The results indicate that pseudocode-based code translation is notably helpful in complementing direct translation for relatively complicated programs.
Our case studies in \Cref{subsubsec:advantages} will further discuss the advantages of pseudocode-based translation, which also align with this observation.

\begin{figure}[t]
\centering
\includegraphics[width=0.986\textwidth]{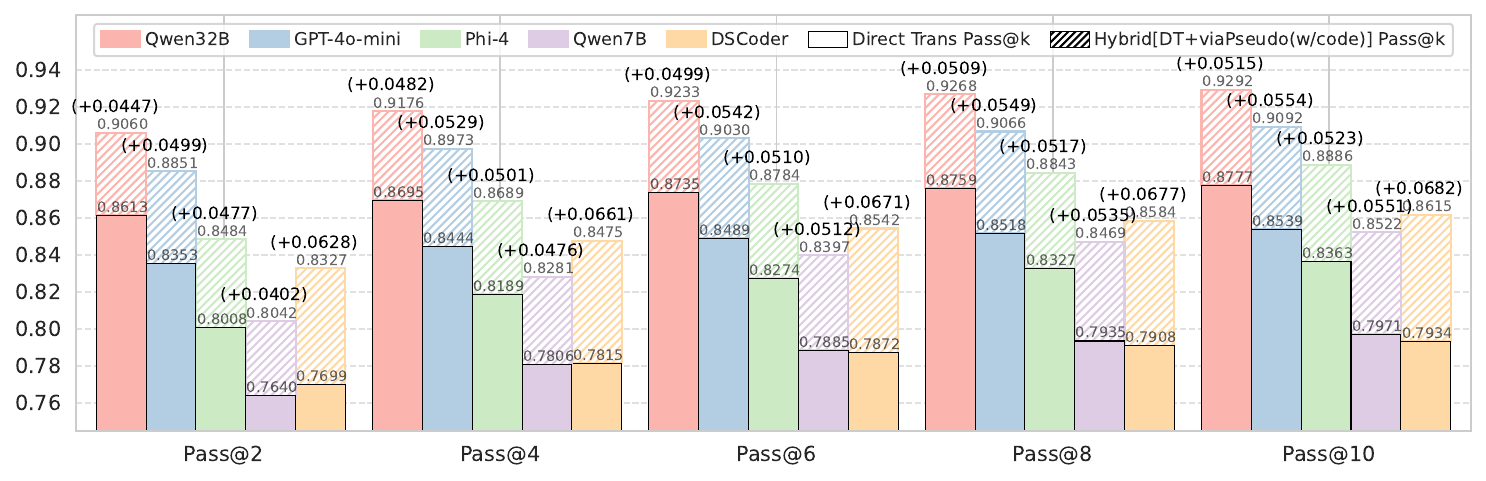}
\setlength{\abovecaptionskip}{0pt}
\caption{Pass Rates of Direct Translation {\sd} and Hybrid Strategy {\sdvpp} with Different Numbers of Attempts}
\label{fig: rq1-passk-all}
\end{figure}

{\textit{Pass Rates with Different Numbers of Attempts.} We also compare the pass rates of direct translation {\sd} and hybrid strategy {\sdvpp} with varying numbers of attempts (i.e., pass@\textit{k}, \textit{k}${\in}$\{2,4,6,8,10\}).
As shown in \Cref{fig: rq1-passk-all}, {\sdvpp} consistently shows higher pass rates than {\sd} under different \textit{k} values. 
In addition, the benefit of using the hybrid strategy (demonstrated by improvement of using {\sdvpp}, indicated by ``(+\textit{value})'' in \Cref{fig: rq1-passk-all}) slightly increases with larger \textit{k} values, indicating that the hybrid strategy can better leverage more attempts to generate correct translations. These further confirm the effectiveness of pseudocode-based translation in complementing direct translation, suggesting the adoption of hybrid strategies with multiple attempts to enhance code translation accuracy. 
}

\begin{summary}
Pseudocode-based code translation can effectively complement the widely adopted direct translation approach to translate programs that cannot be correctly handled by direct translation alone, with a more significant improvement for more complicated programs. 
We recommend that practitioners adopt hybrid strategies that combine the strengths of direct translation and pseudocode-based translation to enhance translation accuracy. 
\end{summary}

\textit{\textbf{Usefulness of Original Programs.}}
The comparison between the pure pseudocode-based strategies with and without the original program (i.e., {\svp} \textit{vs.} {\svpp}) further reveals the need to include the original program in pseudocode-based code translation. 
Specifically, the pure pseudocode-based strategy with the original program as context ({\svpp}) achieved an average pass@10 rate of 0.9270, 0.8357, and 0.6373 on the three levels' solution programs, respectively, while the strategy without the original program ({\svp}) only achieved pass@10 rates of 0.8634, 0.7392, and 0.5191 accordingly. The results suggest that the pseudocode generated by the translator LLM missed or misinterpreted details necessary for the correct generation of programs in the target PL in some cases. We investigate the concrete symptoms of this issue via case studies and discuss them in \Cref{subsec: case-study}.

Meanwhile, the helpfulness of including original programs becomes less obvious for the hybrid strategies (i.e., {\sdvp} vs. {\sdvpp}). Specifically, the pass@10 rates of {\sdvp} and {\sdvpp} are generally comparable (with less than 0.01 difference in pass@10) in most comparisons. This suggests that the original program is less necessary to supplement pseudocode when using hybrid strategies, as direct translation may already provide access to referencing details in the original program for translation tasks that heavily rely on such details, further confirming the complementary strengths of both approaches.
The exceptional cases include GPT-4o-mini on hard-level tasks and Qwen7B on medium-level and hard-level tasks, where the hybrid strategy with the original program as context ({\sdvpp}) achieved obviously higher pass@10 rates. Therefore, it is generally beneficial to include the original program as a backup reference implementation of pseudocode when translating relatively complicated programs.

\begin{summary}
The pseudocode generated by translator LLMs may miss details necessary for generating semantically equivalent programs in the target PL. Including the original program as context is generally suggested to harness the benefits of pseudocode-based code translation, especially for complicated programs.
\end{summary}

\subsection{RQ2: Effectiveness Across Different Source and Target Programming Languages}
\label{subsec:rq2}

Differences in syntax and semantics among PLs may cause varying translation difficulty across PL pairs \cite{icse24translationllmsurvey,icse25intertrans}. In this RQ, we thus further compare the effectiveness of the pseudocode-based translation strategies on different source-target PL pairs. 
The comparison aims to (1) verify whether the helpfulness of pseudocode-based translation identified in RQ1 persists across different source and target PLs, and (2) explore if the pseudocode-based translation leads to significant improvements for certain PL pairs.
The empirical findings are intended to guide developers in adopting pseudocode-based approaches for code translation involving different PLs.
In this RQ, we focus on the improvement of the hybrid translation strategy with context ({\sdvpp}), which showed generally optimal performance in RQ1, relative to the conventional direct translation strategy ({\sd}).

\begin{figure}[t]
\captionsetup[sub]{skip=0.618pt}
\centering

\begin{subfigure}{0.328\textwidth}
\includegraphics[width=\textwidth]{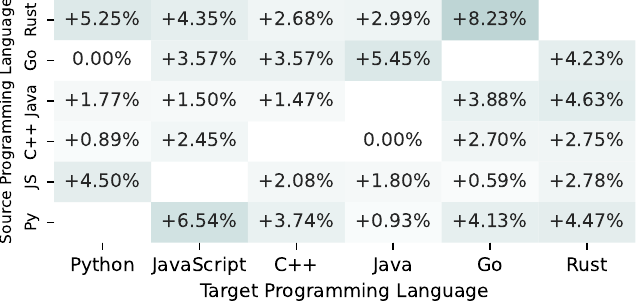}
\subcaption[]{Qwen32B (Easy Tasks)}
\end{subfigure}
\begin{subfigure}{0.328\textwidth}
\includegraphics[width=\textwidth]{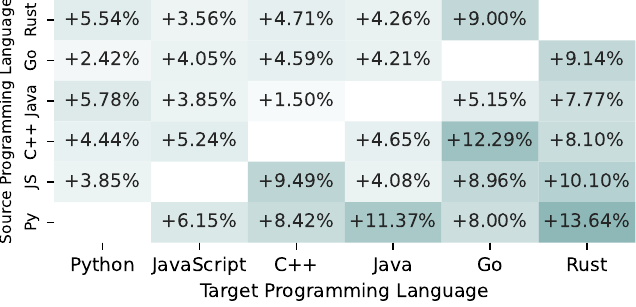}
\subcaption[]{Qwen32B (Med Tasks)}
\end{subfigure}
\begin{subfigure}{0.328\textwidth}
\includegraphics[width=\textwidth]{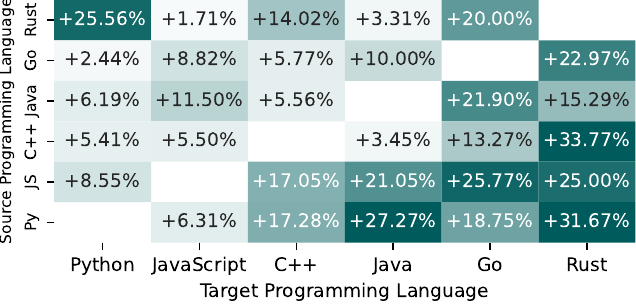}
\subcaption[]{Qwen32B (Hard Tasks)}
\end{subfigure}

\begin{subfigure}{0.328\textwidth}
\includegraphics[width=\textwidth]{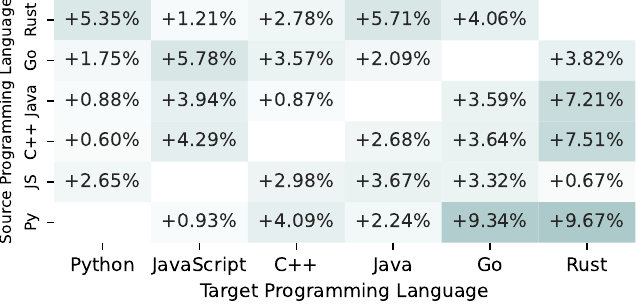}
\subcaption[]{Phi-4 (Easy Tasks)}
\end{subfigure}
\begin{subfigure}{0.328\textwidth}
\includegraphics[width=\textwidth]{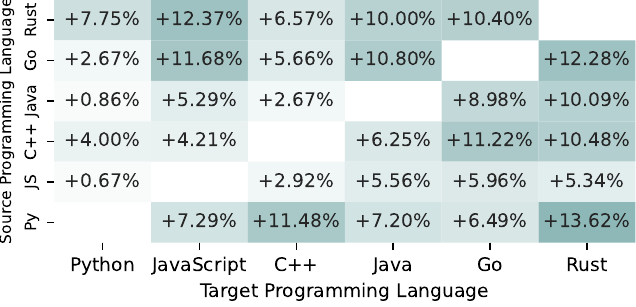}
\subcaption[]{Phi-4 (Med Tasks)}
\end{subfigure}
\begin{subfigure}{0.328\textwidth}
\includegraphics[width=\textwidth]{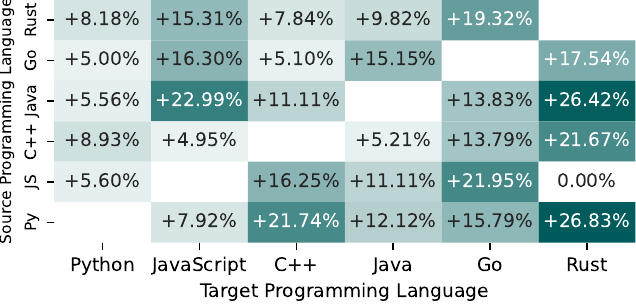}
\subcaption[]{Phi-4 (Hard Tasks)}
\end{subfigure}

\begin{subfigure}{0.328\textwidth}
\includegraphics[width=\textwidth]{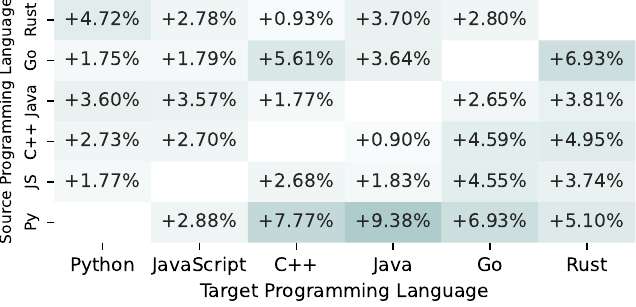}
\subcaption[]{GPT-4o-mini (Easy Tasks)}
\end{subfigure}
\begin{subfigure}{0.328\textwidth}
\includegraphics[width=\textwidth]{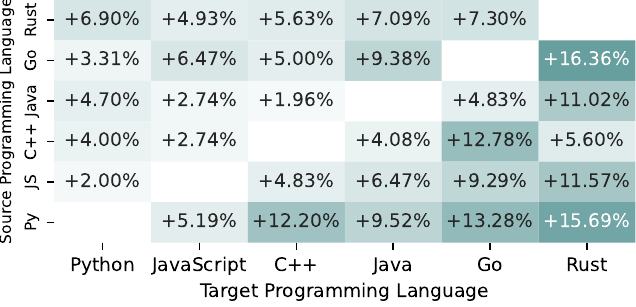}
\subcaption[]{GPT-4o-mini (Med Tasks)}
\end{subfigure}
\begin{subfigure}{0.328\textwidth}
\includegraphics[width=\textwidth]{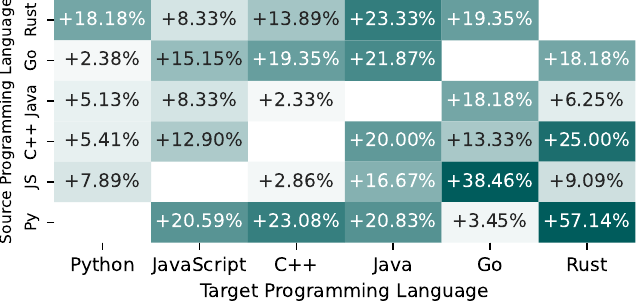}
\subcaption[]{GPT-4o-mini (Hard Tasks)}
\end{subfigure}

\begin{subfigure}{0.328\textwidth}
\includegraphics[width=\textwidth]{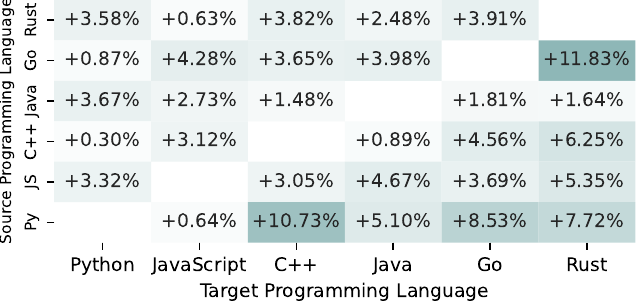}
\subcaption[]{Qwen7B (Easy Tasks)}
\end{subfigure}
\begin{subfigure}{0.328\textwidth}
\includegraphics[width=\textwidth]{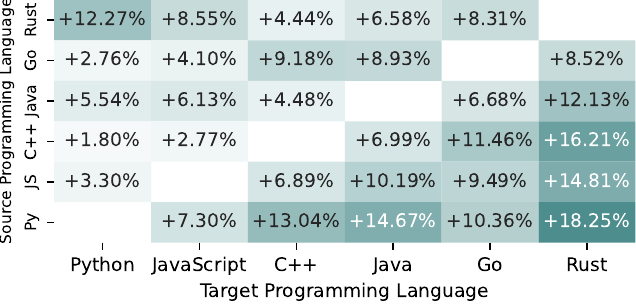}
\subcaption[]{Qwen7B (Med Tasks)}
\end{subfigure}
\begin{subfigure}{0.328\textwidth}
\includegraphics[width=\textwidth]{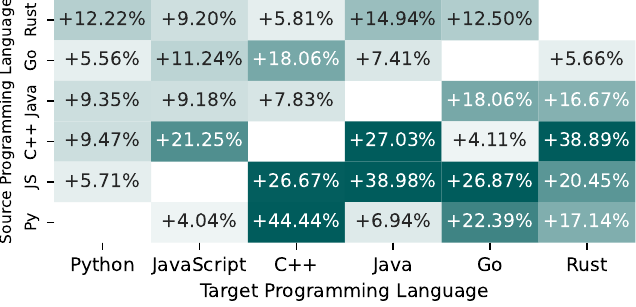}
\subcaption[]{Qwen7B (Hard Tasks)}
\end{subfigure}

\begin{subfigure}{0.328\textwidth}
\includegraphics[width=\textwidth]{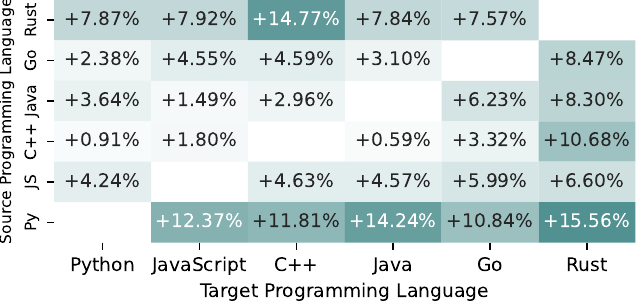}
\subcaption[]{DSCoder (Easy Tasks)}
\end{subfigure}
\begin{subfigure}{0.328\textwidth}
\includegraphics[width=\textwidth]{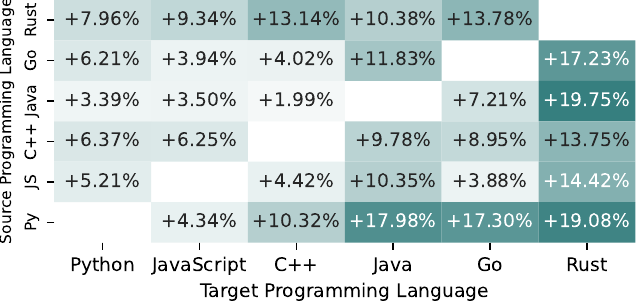}
\subcaption[]{DSCoder (Med Tasks)}
\end{subfigure}
\begin{subfigure}{0.328\textwidth}
\includegraphics[width=\textwidth]{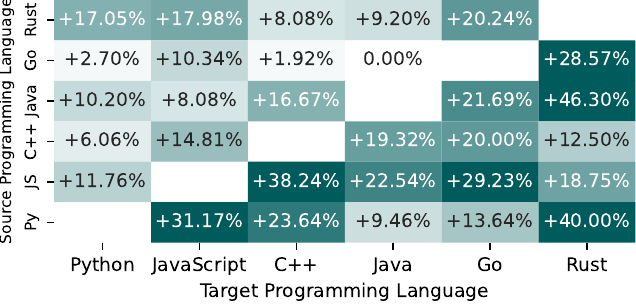}
\subcaption[]{DSCoder (Hard Tasks)}
\end{subfigure}

\caption{Relative Improvement in Pass@10 Using Hybrid Strategy {\sdvpp} Compared with Direct Translation Strategy {\sd} Across Different Source-Target Programming Languages.
(The \colorbox[HTML]{599494}{\color{white}greener}, the more improvement.)\label{fig: rq2combinewcodepassimprove}}
\end{figure}

\textbf{\textit{Overall Effectiveness.}}
\Cref{fig: rq2combinewcodepassimprove} presents the relative improvement in pass@10 rates of the hybrid strategy {\sdvpp} over the baseline {\sd} across different PL pairs. Each row/column in a heatmap corresponds to a specific source/target PL.
In general, we observed that the translation accuracy between almost all PL pairs improves (as indicated by positive values in figures) when using the {\sdvpp} strategy compared to {\sd}. Across all 450 PL-pair-model-difficulty combinations, 147 (32.7\%) combinations achieve even more than 10\% improvement. 
The improvement is consistently observed across different LLMs and task difficulties, with more significant improvement on harder tasks and weaker LLMs, similar to trends observed in RQ1.
The results confirm the generalizable benefits of adopting pseudocode-based translation among different PL pairs.

\textbf{\textit{Comparison Among PL Pairs.}}
As shown by the different colors in heatmaps within \Cref{fig: rq2combinewcodepassimprove}, the improvement varies across different PL pairs, which may be due to the differences across PL pairs and the varying training corpus of different PLs. 
Nevertheless, we observe several consistent trends across different LLMs and task difficulties. 
Specifically, we first noticed that the improvement of code translation \textit{from flexible and lightweight PLs (e.g., Python and JavaScript) to stricter and more complex PLs (e.g., Rust and Go)} is often more significant than the improvement in the opposite direction, as shown by more dark green colors near the right-bottom corner of each heatmap. 
For example, when conducting Python-to-Rust translation and JavaScript-to-Go translation on medium-level tasks, the improvement on Qwen32B is 13.64\% and 8.96\%, respectively; in comparison, it is only 5.54\% and 4.05\% in the opposite direction, respectively. 
We conjecture this is because transitive translation based on pseudocode helps guide LLMs to generate code fitting the requirements enforced in stricter and more complex PLs (e.g., Python and JavaScript are flexible with dynamic typing and rich syntax sugar, while C++ and Java require static type declaration and more rigid syntax, and Go and Rust further enforce more rules on variable usage and memory management).

In addition, we observed that code translation \textit{to Rust} (rightmost column in heatmaps) generally gains significant improvement, regardless of the source PL. 
We conjecture this may also result from the relatively lower resource of the Rust training corpus \cite{tmlr25crosslingual}, particularly parallel code pairs between Rust and other PLs, which are vital to code-to-code translation \cite{avataracl} while limited in the practical training corpus.
In comparison, there may be more natural language-Rust code pairs, such as Rust documentation and programming tutorials, which help LLMs learn to generate Rust code matching natural language descriptions. The pseudocode-based translation enables LLMs to generate Rust code in this manner, which may better align with the training corpus. 
Also, translating \textit{from Rust} (top row in heatmaps) gains noticeable improvement in many trials, suggesting that pseudocode also helps the understanding of original programs written in Rust.
The results demonstrate the helpfulness of pseudocode in assisting code translation involving a low-training-resource PL.

\begin{summary}
Pseudocode-based translation generally improves code translation accuracy across most PL pairs.
In addition, improvement is more significant when translating from flexible and lightweight PLs to strict and complex PLs, and when dealing with the low-training-resource Rust. Practitioners are highly advised to consider pseudocode-based translation in these scenarios.
\end{summary}

\subsection{RQ3: Comparison with Programming Language as Transitive Intermediary}
\label{subsec:rq3}

A recent study \cite{icse25intertrans} reveals that PLs themselves can also serve as effective intermediaries in transitive code translation to complement direct translation. 
For example, translating Python to Rust and then to Java helps resolve some translation tasks that cannot be handled by direct Python-to-Java translation~\cite{icse25intertrans}. 
Given PL as an effective baseline of transitive translation, this RQ verifies whether pseudocode can serve as a more general and effective transitive intermediary. 

To conduct this comparison, we follow \citet{icse25intertrans} to implement a baseline by substituting the transitive intermediary in our hybrid translation strategy (i.e., pseudocode) with a specific PL $L_i$, which is denoted as $\mathcal{D}\&\mathcal{PL}_{L_i}$. 
Specifically, given a source PL $L_s$, a target PL $L_t$, and an intermediate PL $L_i$ ($L_i \notin \{L_s, L_t\}$), the baseline first translates the original program into an intermediate program in $L_i$, and then translates this intermediate program into $L_t$. The resulting program in $L_t$ is taken as the final translated result. 
The translations are performed by LLMs using the same prompt shown in \Cref{subfig: prompt-directtranslate}. 
Since no universally effective intermediate PL has been identified \cite{icse25intertrans}, we implement the baseline for each studied PL and compare our {\sdvpp} strategy against each $\mathcal{D}\&\mathcal{PL}_{L_i}$, where $L_i \in \left\{\text{Python}, \text{JavaScript}, \text{C++}, \text{Java}, \text{Go}, \text{Rust}\right\}$, with pass@10 as the metric.

\begin{figure}[t]
\captionsetup[sub]{skip=0.309pt}
\centering

\begin{subfigure}{0.328\textwidth}
\includegraphics[width=\textwidth]{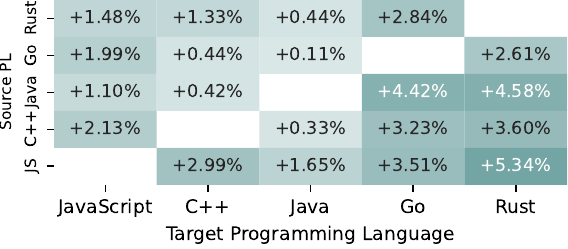}
\subcaption[]{\textit{vs. $\mathcal{D}\&\mathcal{PL}_{\text{Python}}$} (Qwen32B)}
\end{subfigure}
\begin{subfigure}{0.328\textwidth}
\includegraphics[width=\textwidth]{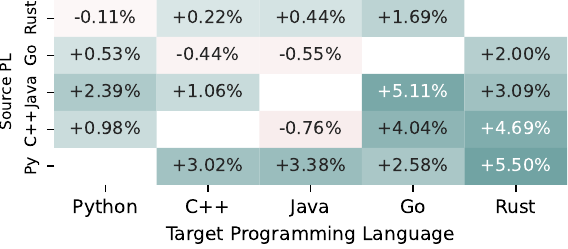}
\subcaption[]{\textit{vs. $\mathcal{D}\&\mathcal{PL}_{\text{JavaScript}}$} (Qwen32B)}
\end{subfigure}
\begin{subfigure}{0.328\textwidth}
\includegraphics[width=\textwidth]{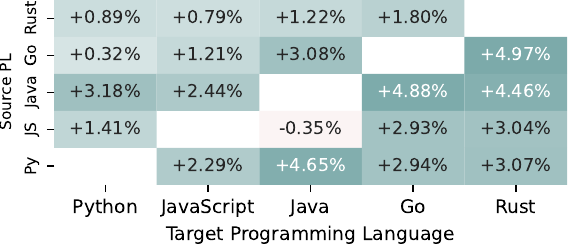}
\subcaption[]{\textit{vs. $\mathcal{D}\&\mathcal{PL}_{\text{C++}}$} (Qwen32B)}
\end{subfigure}

\begin{subfigure}{0.328\textwidth}
\includegraphics[width=\textwidth]{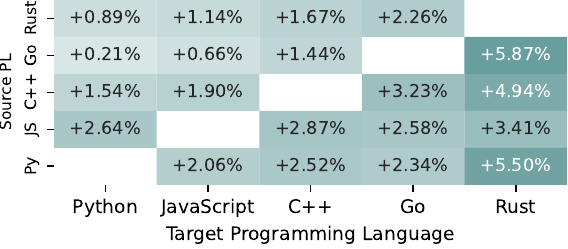}
\subcaption[]{\textit{vs. $\mathcal{D}\&\mathcal{PL}_{\text{Java}}$} (Qwen32B)}
\end{subfigure}
\begin{subfigure}{0.328\textwidth}
\includegraphics[width=\textwidth]{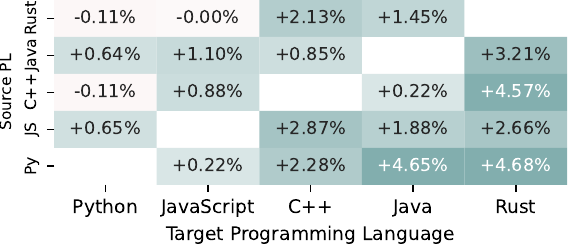}
\subcaption[]{\textit{vs. $\mathcal{D}\&\mathcal{PL}_{\text{Go}}$} (Qwen32B)}
\end{subfigure}
\begin{subfigure}{0.328\textwidth}
\includegraphics[width=\textwidth]{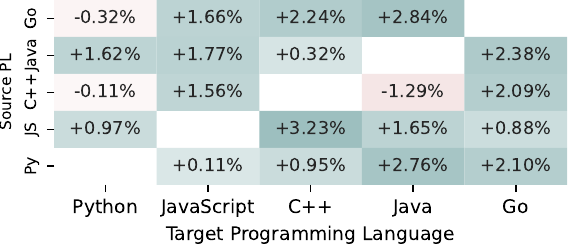}
\subcaption[]{\textit{vs. $\mathcal{D}\&\mathcal{PL}_{\text{Rust}}$} (Qwen32B)}
\end{subfigure}

\caption{Relative Improvement in Pass@10 Rates of Qwen32B Using Hybrid Strategy Based on Pseudocode Compared to Using Hybrid Strategy Based on an Intermediate PL. 
(The \colorbox[HTML]{599494}{\color{white}greener}, the more improvement. \colorbox[HTML]{E8C6C6}{\color{black}Red value} means performance drop.)\label{fig: rq3combinewcodepassimprovevscombinepl-twomodel}}
\end{figure}

\Cref{fig: rq3combinewcodepassimprovevscombinepl-twomodel} presents the comparison results of pseudocode-based translation ({\sdvpp}) versus intermediate PL-based translation ($\mathcal{D}\&\mathcal{PL}_{L_i}$) on Qwen32B. 
The results show that pseudocode-based translations achieve higher code translation accuracy than intermediate PL-based translations in most situations.
Specifically, among all 120 comparisons ($5$ source PLs $\times$ $4$ target PLs $\times$ $6$ transitive PLs), pseudocode-based translation wins in 109 comparisons (green cells), whereas it loses in only 11 comparisons (red cells) with marginal differences (less than 1\%).
Although certain intermediate PLs are more helpful than pseudocode for some source-target PL pairs (e.g., using JavaScript as the intermediate PL for Rust-to-Python, Go-to-C++, Go-to-Java, and C++-to-Java translation, \Cref{fig: rq3combinewcodepassimprovevscombinepl-twomodel}(b)), the pseudocode-based translations yield a generally more pronounced outperformance.
The observed trends and conclusions are consistent across the other studied LLMs (i.e., Qwen7B, Phi-4, GPT-4o-mini, and DSCoder). To avoid redundancy, we omit their detailed results here. These results are available in our artifact~\cite{artifact}.

Through case studies, we found that the advantage of pseudocode over PLs is its role as a \emph{general abstraction} to bridge differences across diverse source-target PL pairs.
Specifically, aligning with findings in \cite{icse25intertrans}, we observed that a transitive PL is effective when it bridges the different features and APIs between the source and target PLs; meanwhile, different PL pairs may require varying transitive PLs. An inappropriate choice of intermediate PL may even increase the gap between the source and target PLs and introduce extra burdens in code translation, hindering the translation accuracy. In addition, an intermediate PL may also bring LLMs new burdens to satisfy its syntax and semantics, complicating the translation process.
In contrast, flexible and descriptive pseudocode presents code intent and logic, offering a PL-agnostic abstraction that guides LLMs to focus on semantics instead of detailed PL-specific implementations, thereby offering better effectiveness in general.
We also examined cases where pseudocode underperformed PLs in transitive translation and found that inferiority mainly stemmed from missing essential details in the pseudocode (which will be discussed in \Cref{subsubsec:limitations}, Limit-2). In comparison, programs in effective intermediate PLs often preserve the complete code semantics, facilitating the generation of a semantically preserving translation result.
These findings highlight the advantages of pseudocode as a general effective intermediary, as well as the need for enhancing its quality to fully harness its effectiveness.

\begin{summary}
Pseudocode serves as a more general and effective intermediary than a specific PL in transitive code translation for most studied PL pairs. This is because pseudocode provides a PL-agnostic abstraction of code semantics that generally bridges differences across diverse source-target PL pairs, as an effective transitive PL does for suitable PL pairs. In development, practitioners can prioritize pseudocode as a universal bridge for translation across various PLs.
\end{summary}

\subsection{RQ4: Effectiveness of Code Translation based on High-Quality Pseudocode}
\label{subsec:rq4}

The pseudocode generated by the studied LLMs may introduce noise (e.g., incorrect or ambiguous descriptions of code intent and logic, which will be discussed in \Cref{subsec: case-study}). 
Such low-quality pseudocode can mislead code translation. This RQ therefore examines whether better performance can be achieved when high-quality pseudocode is available, as well as identifies the bottlenecks in pseudocode-based code translation.
Specifically, DeepSeek-R1 \cite{r1techreport} is found capable of helping annotate human-written-like pseudocode with high accuracy and naturalness for concrete programs \cite{pseudoeval}. Therefore, in this RQ, we take DeepSeek-R1 as a high-quality pseudocode generator\footnote{To avoid potential data contamination of DeepSeek-R1, whose knowledge cutoff date is unknown, we only adopt it to prepare high-quality pseudocode following \citet{pseudoeval}. We do not include it as a translator LLM to study.} and evaluate the translation accuracy of the studied LLMs based on DeepSeek-R1-generated high-quality pseudocode. 
The investigation results help indicate the potential effectiveness of pseudocode-based translation when higher-quality pseudocode is available (e.g., written by humans or generated by enhanced pseudocode generation methods).

\begin{table}[t]
    \caption{Pass@10 of Round-Trip Code Regeneration Based on Pseudocode Generated by
    Translator LLM (Self-Gen) and DeepSeek-R1 (R1-Gen).\label{table: rq4roundtripvalidate}}
    \setlength{\tabcolsep}{7.5pt}
    \footnotesize
    \begin{tabular}{ccc@{\hskip 0pt}ccc@{\hskip 0pt}ccc@{\hskip 0pt}c}
        \toprule 
\multirow{2}{*}{\textbf{LLM}} & \multicolumn{3}{c}{\textbf{Easy Tasks}}                 & \multicolumn{3}{c}{\textbf{Medium Tasks}}               & \multicolumn{3}{c}{\textbf{Hard Tasks}}                 \\
                              & \textbf{LLM-Gen}  & \multicolumn{2}{c}{\textbf{R1-Gen}} & \textbf{LLM-Gen}  & \multicolumn{2}{c}{\textbf{R1-Gen}} & \textbf{LLM-Gen}  & \multicolumn{2}{c}{\textbf{R1-Gen}} \\ \midrule
Qwen32B                       & 0.9516            & 0.9670         & {\scriptsize (+1.62\%)}          & 0.8976            & 0.9042         & {\scriptsize (+0.74\%)}          & 0.7352            & 0.8050         & {\scriptsize (+9.49\%)}          \\
Phi-4                         & 0.8889            & 0.9425         & {\scriptsize (+6.03\%)}          & 0.8010            & 0.8365         & {\scriptsize (+4.43\%)}          & 0.6738            & 0.6773         & {\scriptsize (+0.52\%)}          \\
GPT-4o-mini                   & 0.8865            & 0.9583         & {\scriptsize (+8.10\%)}          & 0.7583            & 0.8646         & {\scriptsize (+14.02\%)}         & 0.5532            & 0.6986         & {\scriptsize (+26.28\%)}         \\
Qwen7B                        & 0.8008            & 0.9267         & {\scriptsize (+15.72\%)}         & 0.6413            & 0.8156         & {\scriptsize (+27.18\%)}         & 0.3865            & 0.5567         & {\scriptsize (+44.04\%)}         \\
DS-Coder                      & 0.8774            & 0.9353         & {\scriptsize (+6.60\%)}          & 0.7833            & 0.8188         & {\scriptsize (+4.53\%)}          & 0.5875            & 0.6525         & {\scriptsize (+11.06\%)}         \\ 
        \bottomrule \multicolumn{10}{l}{\scriptsize \textit{(+x.xx\%): performance improvement of a strategy based on R1-generated pseudocode compared to translator-LLM-generated pseudocode.}}
    \end{tabular}
\end{table}

{
\textit{Quality Check of R1-Generated Pseudocode:} Before comparing the code translation accuracy based on different pseudocode, we verify the assumed higher quality of pseudocode generated by DeepSeek-R1 over the studied LLMs using round-trip validation, an effective method to validating quality of semantic representations like code summaries \cite{codesumroundtrip}, natural language translations \cite{roundtripnlp}, and pseudocode \cite{pseudoeval}. Specifically, we use each studied LLM to regenerate the original program based on the pseudocode generated by itself and DeepSeek-R1, respectively, following the prompt in \Cref{subfig: prompt-pseudocodewocode} with the original PL. The regeneration pass rates can indicate the pseudocode quality, since higher-quality pseudocode should better describe code intent and logic of original programs, thus facilitating regeneration. As shown in \Cref{table: rq4roundtripvalidate}, the pass rates based on R1-generated pseudocode are consistently higher than translator-generated pseudocode, suggesting the higher quality of R1-generated pseudocode in describing the code intent and logic of original programs.
}

\textit{\textbf{Improvement with High-quality Pseudocode.}} 
\Cref{table: rq4qualitycombinewcode} compares the pass@10 rates of different translation strategies based on high-quality pseudocode generated by DeepSeek-R1, and pseudocode generated by the studied LLMs themselves (i.e., the default setup studied in previous RQs), respectively.
The results demonstrate that the code translation accuracy consistently improves across LLMs and strategies after switching to DeepSeek-R1-generated pseudocode.
Among all strategies, the pure pseudocode-based strategy ({\svp}) benefited the most, with average improvements of 8.16\%, 13.45\%, and 22.15\% on three difficulty levels, respectively, mitigating the underperformance relative to direct translation ({\sd}) observed in RQ1. 
These results indicate pseudocode quality as a critical bottleneck for pseudocode-based translation.
In addition, we observed that hybrid strategies ({\sdvp} and {\sdvpp}) continuously outperformed single-approach strategies ({\sd}, {\svp}, {\svpp}), indicating that combining direct and pseudocode-based translation remains beneficial even with higher-quality pseudocode. 
With DeepSeek-R1-generated high-quality pseudocode, the best pass@10 scores of all LLMs with optimal strategy improved to 0.9646-0.9835, 0.8861--0.9512, and 0.6747--0.8286 across three difficulty levels, demonstrating the promising potential of pseudocode-based code translation.

\begin{table}[t]
    \caption{Pass@10 of Code Translation Based on Pseudocode Generated by
    Translator LLM (Self-Gen) and DeepSeek-R1 (R1-Gen).\label{table: rq4qualitycombinewcode}}
    \setlength{\tabcolsep}{4pt}
    \footnotesize
    \begin{tabular}{crcc@{\hskip 0pt}ccc@{\hskip 0pt}ccc@{\hskip 0pt}c}
        \toprule \multirow{2}{*}{\textbf{LLM}}                                                                                                                                              & \multirow{2}{*}{\textbf{Strategy}} & \multicolumn{3}{c}{\textbf{Easy Tasks}} & \multicolumn{3}{c}{\textbf{Medium Tasks}} & \multicolumn{3}{c}{\textbf{Hard Tasks}} \\
                                                                                                                                                                                            &                                    & \textbf{Self-Gen}                       & \multicolumn{2}{c}{\textbf{R1-Gen}} & \textbf{Self-Gen}                      & \multicolumn{2}{c}{\textbf{R1-Gen}} & \textbf{Self-Gen} & \multicolumn{2}{c}{\textbf{R1-Gen}} \\
        \midrule \multirow{5}{*}{Qwen32B}                                                                                                                                                   & {\sd}                              & 0.9480                                  & -                                         & -                                      & 0.8772                                    & -                 & -                                        & 0.7059 & -               & -                                \\
                                                                                                                                                                                            & {\svp}                             & 0.9442                                  & 0.9569                                    & \textbf{{\scriptsize(+1.35\%)}}        & 0.8736                                    & 0.8971            & \textbf{{\scriptsize(+2.69\%)}}          & 0.6768 & 0.7376          & \textbf{{\scriptsize(+8.98\%)}}  \\
                                                                                                                                                                                            & {\svpp}                            & 0.9648                                  & 0.9718                                    & {\scriptsize(+0.73\%)}                 & 0.9001                                    & 0.9227            & {\scriptsize(+2.51\%)}                   & 0.7553 & 0.7858          & {\scriptsize(+4.04\%)}           \\
                                                                                                                                                                                            & {\sdvp}                            & 0.9773                                  & 0.9833                                    & {\scriptsize(+0.61\%)}                 & 0.9403                                    & \textbf{0.9512}   & {\scriptsize(+1.16\%)}                   & 0.7948 & 0.8260          & {\scriptsize(+3.93\%)}           \\
                                                                                                                                                                                            & {\sdvpp}                           & 0.9773                                  & \textbf{0.9835}                           & {\scriptsize(+0.63\%)}                 & 0.9327                                    & 0.9476            & {\scriptsize(+1.60\%)}                   & 0.7988 & \textbf{0.8286} & {\scriptsize(+3.73\%)}           \\
        \midrule \multirow{5}{*}{Phi-4}                                                                                                                                                     & {\sd}                              & 0.9276                                  & -                                         & -                                      & 0.8313                                    & -                 & -                                        & 0.6281 & -               & -                                \\
                                                                                                                                                                                            & {\svp}                             & 0.8782                                  & 0.9287                                    & \textbf{{\scriptsize(+5.75\%)}}        & 0.7744                                    & 0.8402            & \textbf{{\scriptsize(+8.50\%)}}          & 0.5832 & 0.6433          & \textbf{{\scriptsize(+10.31\%)}} \\
                                                                                                                                                                                            & {\svpp}                            & 0.9254                                  & 0.9572                                    & {\scriptsize(+3.44\%)}                 & 0.8347                                    & 0.8796            & {\scriptsize(+5.38\%)}                   & 0.6463 & 0.6986          & {\scriptsize(+8.09\%)}           \\
                                                                                                                                                                                            & {\sdvp}                            & 0.9646                                  & \textbf{0.9781}                           & {\scriptsize(+1.40\%)}                 & 0.8947                                    & \textbf{0.9192}   & {\scriptsize(+2.74\%)}                   & 0.7083 & \textbf{0.7563} & {\scriptsize(+6.78\%)}           \\
                                                                                                                                                                                            & {\sdvpp}                           & 0.9610                                  & 0.9759                                    & {\scriptsize(+1.55\%)}                 & 0.8903                                    & 0.9127            & {\scriptsize(+2.52\%)}                   & 0.7040 & 0.7385          & {\scriptsize(+4.90\%)}           \\
        \midrule \multirow{5}{*}{\begin{tabular}[c]{@{}c@{}}GPT-4o\\ -mini\end{tabular}}                                                                                                    & {\sd}                              & 0.9279                                  & -                                         & -                                      & 0.8552                                    & -                 & -                                        & 0.6667 & -               & -                                \\
                                                                                                                                                                                            & {\svp}                             & 0.8647                                  & 0.9448                                    & \textbf{{\scriptsize(+9.26\%)}}        & 0.7219                                    & 0.8538            & \textbf{{\scriptsize(+18.27\%)}}         & 0.5000 & 0.6752          & \textbf{{\scriptsize(+35.04\%)}} \\
                                                                                                                                                                                            & {\svpp}                            & 0.9239                                  & 0.9641                                    & {\scriptsize(+4.35\%)}                 & 0.8635                                    & 0.8973            & {\scriptsize(+3.91\%)}                   & 0.6837 & 0.7333          & {\scriptsize(+7.25\%)}           \\
                                                                                                                                                                                            & {\sdvp}                            & 0.9658                                  & 0.9787                                    & {\scriptsize(+1.34\%)}                 & 0.9054                                    & \textbf{0.9338}   & {\scriptsize(+3.14\%)}                   & 0.7404 & \textbf{0.7972} & {\scriptsize(+7.67\%)}           \\
                                                                                                                                                                                            & {\sdvpp}                           & 0.9621                                  & \textbf{0.9790}                           & {\scriptsize(+1.76\%)}                 & 0.9144                                    & 0.9327            & {\scriptsize(+2.00\%)}                   & 0.7617 & 0.7915          & {\scriptsize(+3.91\%)}           \\
        \midrule \multirow{5}{*}{Qwen7B}                                                                                                                                                    & {\sd}                              & 0.9073                                  & -                                         & -                                      & 0.7928                                    & -                 & -                                        & 0.5400 & -               & -                                \\
                                                                                                                                                                                            & {\svp}                             & 0.7735                                  & 0.9216                                    & \textbf{{\scriptsize(+19.15\%)}}       & 0.5928                                    & 0.8042            & \textbf{{\scriptsize(+35.66\%)}}         & 0.3286 & 0.5390          & \textbf{{\scriptsize(+64.03\%)}} \\
                                                                                                                                                                                            & {\svpp}                            & 0.8989                                  & 0.9497                                    & {\scriptsize(+5.65\%)}                 & 0.7929                                    & 0.8575            & {\scriptsize(+8.15\%)}                   & 0.5480 & 0.6426          & {\scriptsize(+17.26\%)}          \\
                                                                                                                                                                                            & {\sdvp}                            & 0.9336                                  & \textbf{0.9646}                           & {\scriptsize(+3.32\%)}                 & 0.8395                                    & 0.8831            & {\scriptsize(+5.19\%)}                   & 0.5849 & 0.6598          & {\scriptsize(+12.81\%)}          \\
                                                                                                                                                                                            & {\sdvpp}                           & 0.9420                                  & 0.9638                                    & {\scriptsize(+2.31\%)}                 & 0.8568                                    & \textbf{0.8861}   & {\scriptsize(+3.42\%)}                   & 0.6154 & \textbf{0.6747} & {\scriptsize(+9.64\%)}           \\
        \midrule \multirow{5}{*}{DSCoder}                                                                                                                                                   & {\sd}                              & 0.8964                                  & -                                         & -                                      & 0.7876                                    & -                 & -                                        & 0.5586 & -               & -                                \\
                                                                                                                                                                                            & {\svp}                             & 0.8562                                  & 0.9170                                    & \textbf{{\scriptsize(+7.10\%)}}        & 0.7335                                    & 0.7979            & \textbf{{\scriptsize(+8.78\%)}}          & 0.5069 & 0.5752          & {\scriptsize(+13.47\%)}          \\
                                                                                                                                                                                            & {\svpp}                            & 0.9218                                  & 0.9509                                    & {\scriptsize(+3.16\%)}                 & 0.7875                                    & 0.8540            & {\scriptsize(+8.44\%)}                   & 0.5530 & 0.6504          & \textbf{{\scriptsize(+17.61\%)}} \\
                                                                                                                                                                                            & {\sdvp}                            & 0.9482                                  & 0.9678                                    & {\scriptsize(+2.07\%)}                 & 0.8569                                    & 0.8916            & {\scriptsize(+4.05\%)}                   & 0.6442 & 0.6934          & {\scriptsize(+7.64\%)}           \\
                                                                                                                                                                                            & {\sdvpp}                           & 0.9537                                  & \textbf{0.9726}                           & {\scriptsize(+1.98\%)}                 & 0.8581                                    & \textbf{0.8954}   & {\scriptsize(+4.35\%)}                   & 0.6459 & \textbf{0.7073} & {\scriptsize(+9.51\%)}           \\
        \midrule \multirow{5}{*}{\textit{(Average)}}                                                                                                                                          & {\sd}                              & 0.9214                                  & -                                         & -                                      & 0.8288                                    & -                 & -                                        & 0.6199 & -               & -                                \\
                                                                                                                                                                                            & {\svp}                             & 0.8634                                  & 0.9338                                    & \textbf{{\scriptsize(+8.16\%)}}        & 0.7392                                    & 0.8386            & \textbf{{\scriptsize(+13.45\%)}}         & 0.5191 & 0.6341          & \textbf{{\scriptsize(+22.15\%)}} \\
                                                                                                                                                                                            & {\svpp}                            & 0.9270                                  & 0.9587                                    & {\scriptsize(+3.43\%)}                 & 0.8357                                    & 0.8822            & {\scriptsize(+5.56\%)}                   & 0.6373 & 0.7021          & {\scriptsize(+10.18\%)}          \\
                                                                                                                                                                                            & {\sdvp}                            & 0.9579                                  & 0.9745                                    & {\scriptsize(+1.73\%)}                 & 0.8874                                    & \textbf{0.9158}   & {\scriptsize(+3.20\%)}                   & 0.6945 & 0.7465          & {\scriptsize(+7.49\%)}           \\
                                                                                                                                                                                            & {\sdvpp}                           & 0.9592                                  & \textbf{0.9750}                           & {\scriptsize(+1.64\%)}                 & 0.8905                                    & 0.9149            & {\scriptsize(+2.74\%)}                   & 0.7052 & \textbf{0.7481} & {\scriptsize(+6.09\%)}           \\
        \bottomrule \multicolumn{11}{l}{\scriptsize \textit{(+x.xx\%): performance improvement of a strategy based on R1-generated pseudocode compared to translator-LLM-generated pseudocode.}}
    \end{tabular}
\end{table}

\begin{summary}
The quality of pseudocode hinders accurate pseudocode-based code translation of the studied LLMs.
Higher-quality pseudocode can consistently improve the code translation accuracy of studied LLMs across translation strategies and task difficulties.
Practitioners may also consider human-in-the-loop to prepare high-quality pseudocode or revise the generated pseudocode, which will help guide LLMs to generate more accurate translation results through pseudocode-based code translation.
\end{summary}

\textit{\textbf{LLM-wise Comparison.}} The performance comparison among LLMs further demonstrates the limitations of LLMs in both code understanding and generation during pseudocode-based code translation. 
Specifically, we observed that the performance improvement brought by DeepSeek-R1-generated pseudocode is more significant for weaker LLMs. For example, the improvement of all strategies on Qwen7B ranges from 2.31\% to 64.03\%, while it is only 0.61\% to 8.98\% on Qwen32B, indicating that higher-quality pseudocode is more beneficial for weaker LLMs. This demonstrates the weakness of weaker LLMs in understanding and interpreting the code intent and logic of original programs.
In addition, with DeepSeek-R1-generated pseudocode, the performance gap between weaker LLMs and more powerful LLMs remains. For example, with R1-generated pseudocode, the pass@10 rates of Qwen7B and DSCoder with {\sdvpp} are only 0.6747 and 0.7073 on hard tasks, respectively, which are much lower than 0.8286 of Qwen32B. The results indicate that even given the same code intent and logic (i.e., pseudocode), the weaker LLMs are still less capable in code translation than the more powerful LLMs, indicating that the code implementation capability also hinders the effectiveness of pseudocode-based code translation on weaker LLMs.

\begin{summary}
Both code understanding capability (indicated by the quality of the interpreted pseudocode) and code implementation capability (based on pseudocode) of the studied LLMs hinder their performance in code translation tasks.
Further advancements in generating high-quality pseudocode from programs and implementing accurate code from pseudocode are necessary to fully unleash the potential of pseudocode-based code translation.
\end{summary}

\subsection{Discussion: Advantages and Limitations of Pseudocode-Based Translation}
\label{subsec: case-study}

After understanding the effectiveness of pseudocode-based code translation from the quantitative results in four RQs, we further conduct case studies to attribute its successes and failures to learn concrete insights for the effective adoption of this approach.
{Specifically, two authors with over five years of programming experience analyzed the results of 200 randomly sampled translation tasks (i.e., 100 tasks where direct translation failed while pseudocode-based translation succeeded, and another 100 tasks where direct translation succeeded while pseudocode-based translation failed\footnote{To study reasons of clear successes/failures rather than randomness, we only analyzed the tasks where over five in ten attempts of a strategy passed all tests (successful tasks) and where none attempt passed all tests (failed tasks).}), to identify the advantages/limitations of pseudocode for successful/failed translations compared with direct translation. These tasks are sampled across all the studied LLMs and PL pairs, {representing a 95\% confidence level with a 10\% margin of error}. {The participants first discussed potential categorizations and criteria of advantages and limitations via a pilot study on 40 sampled tasks.}
Then, they independently annotated each task (with substantial agreement indicated by Cohen's Kappa coefficients~\cite{cohen1960coefficient} over 0.78), and next discussed to reach a consensus on all the annotations. Each task can be annotated with multiple advantages/limitations of pseudocode.}
In addition, based on all the findings in this work, we further discuss several research directions to mitigate the limitations and harness the potential for pseudocode-based code translation.

\subsubsection{Key Advantages of Pseudocode-based Code Translation.}
\label{subsubsec:advantages}
We identified three major advantages of pseudocode-based translation over direct translation. 
One of them facilitates the understanding of the original programs and reduces the burden of an integrated step, and the other two help LLMs handle differences across PLs. They align with the limitations of direct translation identified in prior studies \cite{icse24translationllmsurvey,ase23ctstudysjtu,icse25intertrans}.
Practitioners are encouraged to leverage these advantages to translate the original programs or across PL pairs fitting them.

\begin{enumerate}[leftmargin=3em]
    \item[Pro-1] \textit{Pseudocode helps decouple code understanding and implementation in code translation, formulating easier subproblems for LLMs to solve when translating complicated programs.} 
\end{enumerate}

Some programs are compact, implementing multiple operations in a single dense statement by chaining several steps recursively or sequentially. Such programs may result from PL idioms or developers' programming style \cite{methodchainingjavapythonkoltin}. We observed that LLMs struggle to understand the tangled functionality of such complicated programs and re-implement the functionality in the target PL via one-step direct translation. 
LLMs often miss or mess up steps in such cases. In comparison, a pseudocode-based transitive translation splits the code understanding of the complicated program and the implementation of the code intent and logic into two steps, reducing the burden on LLMs and promoting accuracy in each step.
{This advantage was observed in 34\% of analyzed cases where pseudocode-based translation succeeded while direct translation failed.}

\Cref{fig: rq1-type3} illustrates an example of this case. 
Specifically, the original Python program includes only one statement that recursively conducts four operations (\Cref{lst: rq1-type3-py2js-python}). When translating this compact program via the one-step direct translation, even the powerful Qwen32B missed the row-sorting step in the output JavaScript program (\Cref{lst: rq1-type3-py2js-direct}). In comparison, when asked to describe the code intent and logic using pseudocode, Qwen32B was able to interpret the complicated logic of the compact Python code using pseudocode (\Cref{lst: rq1-type3-py2js-pseudocode}). The pseudocode then promotes the successful implementation of the operation {in} JavaScript (\Cref{lst: rq1-type3-py2js-viapseudocode}). Phi-4 and Qwen7B also suffered from similar issues when conducting direct translation and succeeded in pseudocode-based translation.

\begin{figure}[t]
\centering
\begin{subfigure}{0.49\textwidth}
\begin{lstlisting}[language=Python, caption={Original Python Program to Translate}, label={lst: rq1-type3-py2js-python}]
def matrixSum(self, nums: List[List[int]]) -> int:
    return sum(max(col) for col in zip(*[sorted(row) for row in nums]))
    # organizing row-wise sorting, transposition, max 
    # val extraction, and sum in a single expression.
\end{lstlisting}
\end{subfigure}\;
\begin{subfigure}{0.49\textwidth}
\begin{lstlisting}[language=JavaScript, caption={Pseudocode of Original Python Program}, label={lst: rq1-type3-py2js-pseudocode}]
function matrixSum(nums)
    sorted_rows = sort each row in nums
    transposed_matrix = transpose sorted_rows
    max_values = find the maximum value in each col...
    return sum of max_values
\end{lstlisting}
\end{subfigure}

\begin{subfigure}{0.49\textwidth}
\begin{lstlisting}[language=JavaScript, caption={Direct Translation Result}, label={lst: rq1-type3-py2js-direct}]
var matrixSum = function(nums) {
    return nums[0].map((_, colIndex) => Math.max(
     ...nums.map(row => row[colIndex]))//unsorted rows
    ).reduce((sum, maxVal) => sum + maxVal, 0);
};
\end{lstlisting}
\end{subfigure}\;
\begin{subfigure}{0.49\textwidth}
\begin{lstlisting}[language=JavaScript, caption={Pseudocode-based Translation Result}, label={lst: rq1-type3-py2js-viapseudocode}]
var matrixSum = function(nums) {
    const sortedRows = nums.map(row => row.sort((a, b) => a - b)); // sort row before other ops
    const transposedMatrix = sortedRows[0].map(...
    ...
\end{lstlisting}
\end{subfigure}
\setlength{\abovecaptionskip}{-2pt}
\caption{An Example of Python-to-JavaScript Translation Involving Compact Original Programs (Qwen32B)\protect\footref{fn:rq1-example-footnote}}
\label{fig: rq1-type3}
\end{figure}

\begin{enumerate}[leftmargin=3em]
    \item[Pro-2] \textit{Pseudocode abstracts the PL-specific features (e.g., variable scoping, closure capture mode, and syntax sugar), avoids LLMs in following the incompatible practices in original programs.} 
\end{enumerate}

We observed that LLM-based direct translation may overlook the differences in the features of source and target PLs, resulting in incompatible implementations in the translation result that violate the mechanism of the target PL.
Meanwhile, pseudocode describes the original program's intent and logic, hiding PL-specific implementation. When generating target code from pseudocode, LLMs often properly implement the intended function in a manner fitting the mechanism of the target PL, and are less likely to be distracted by the incompatible patterns of source PL in original programs.
{This advantage was observed in 56\% of analyzed cases where pseudocode-based translation outperformed direct translation.}

\Cref{fig: rq1-type1} shows an example of this case with a Go-to-Rust translation. 
Specifically, when translating a Go program with an inner function (\Cref{lst: rq1-type1-go2rust-go}) into Rust via direct translation, Qwen7B mechanically followed the Go program to declare a common inner function in Rust without realizing the difference in default closure capture modes in Go and Rust and failing to indicate a mutable closure capture mode in Rust (\Cref{lst: rq1-type1-go2rust-direct}). 
Meanwhile, based on pseudocode that mainly indicates the code intent and logic (\Cref{lst: rq1-type1-go2rust-pseudocode}), Qwen7B correctly implemented the functionality in a proper manner that fits the memory management of Rust (\Cref{lst: rq1-type1-go2rust-viapseudocode}). Notably, all five studied LLMs failed with the same fault in direct translation while succeeding with the help of pseudocode. 
This also echoes the helpfulness of pseudocode-based code translation in translating from flexible PLs to more rigid PLs observed in RQ2 (\Cref{subsec:rq2}).

\begin{figure}[t]
\centering
\begin{subfigure}{0.49\textwidth}
\begin{lstlisting}[language=Go, caption={Original Go Program to Translate}, label={lst: rq1-type1-go2rust-go}]
func canMakeSquare(grid [][]byte) bool {
    ...
    var res int
    calcRes := func (r byte) { // mutable closure
        if r == 'W' { res++; } //capture by default
    }
    ...
\end{lstlisting}
\end{subfigure}\;
\begin{subfigure}{0.49\textwidth}
\begin{lstlisting}[language=JavaScript, caption={Pseudocode of Original Go Program}, label={lst: rq1-type1-go2rust-pseudocode}]
function canMakeSquare(grid):
    ...
    count = count 'W' in the 2x2 block
    ...



\end{lstlisting}
\end{subfigure}

\begin{subfigure}{0.49\textwidth}
\begin{lstlisting}[language=Rust, caption={Direct Translation Result}, label={lst: rq1-type1-go2rust-direct}]
pub fn can_make_square(grid: Vec<Vec<char>>) -> bool {
    ...
    let mut res = 0;
    let calc_res = |r: char| { //mimic Go code, but
        if r == 'W' {//by default immutable closure
            res += 1;//, thus illegal to mutate res
        }};
    ...
\end{lstlisting}
\end{subfigure}\;
\begin{subfigure}{0.49\textwidth}
\begin{lstlisting}[language=Rust, caption={Pseudocode-based Translation Result}, label={lst: rq1-type1-go2rust-viapseudocode}]
pub fn can_make_square(grid: Vec<Vec<char>>) -> bool {
    ...
    let mut count = 0;
    // flexibly count without involving closure
    for x in 0..2 { for y in 0..2 {
        if grid[i + x][j + y] == 'W' {
            count += 1;
    ...
\end{lstlisting}
\end{subfigure}
\setlength{\abovecaptionskip}{-2pt}
\caption{An Example of Go-to-Rust Translation Involving Unique Closure Capturing in Go and Rust (Qwen7B)\protect\footnotemark}
\label{fig: rq1-type1}
\end{figure}
\footnotetext{\label{fn:rq1-example-footnote}``...'' represents code less related to the illustration and omitted for saving space. The comment is not generated by LLMs but annotated by authors for case explanation.}

\begin{enumerate}[leftmargin=3em]
    \item[Pro-3] \textit{Pseudocode describes the functionality implemented by unique APIs in source PL, allowing LLMs to generate flexible implementations to realize the intended functionality in target PL.}
\end{enumerate}

We observed that LLMs could hallucinate incorrect implementations that fail to realize the functionality of the PL-specific APIs in original programs during direct translation. LLMs struggle to identify or implement a semantically equivalent API counterpart in the target PL based on only the original program, and may even hallucinate non-existent APIs. Meanwhile, pseudocode often describes the functionality implemented with PL-specific APIs in natural language, guiding LLMs to figure out appropriate APIs and API usages in the target PL as they work out a code generation task without being misled by concrete API usages in original programs. 
{This advantage was observed in 33\% of analyzed cases where pseudocode-based translation outperformed direct translation.}

\Cref{fig: rq1-type2} illustrates an example of this case. Specifically, the original Python program uses an API \code{bit\_count} to calculate the number of ones in the binary expression of an integer (\Cref{lst: rq1-type2-py2go-python}). Due to no \code{bit\_count} API in Go, when translating the Python program into Go via direct translation, Phi-4 implemented bit counting with a series of bit operations (\Cref{lst: rq1-type2-py2go-direct}). However, the generated bit operations do not align with the functionality of \code{bit\_count}. Similarly, GPT-4o-mini hallucinated incorrect bit operations, and DSCoder even hallucinated a non-existent API. Instead, the pseudocode generated by Phi-4 clarified the functionality implemented by \code{bit\_count} in Python (\Cref{lst: rq1-type2-py2go-pseudocode}), which guided Phi-4 to generate a correct subroutine \code{countBits} in Go to realize the functionality of \code{bit\_count} (\Cref{lst: rq1-type2-py2go-viapseudocode}). DSCoder and GPT-4o-mini also generated a correct translation based on the pseudocode generated by themselves.

\begin{figure}[t]
\centering
\begin{subfigure}{0.49\textwidth}
\begin{lstlisting}[language=Python, caption={Original Python Program to Translate}, label={lst: rq1-type2-py2go-python}]
def kthCharacter(self, k: int) -> str:
    ...
    return ascii_lowercase[(k-1).bit_count()]



\end{lstlisting}
\end{subfigure}\;
\begin{subfigure}{0.49\textwidth}
\begin{lstlisting}[language=JavaScript, caption={Pseudocode of Original Python Program}, label={lst: rq1-type2-py2go-pseudocode}]
function kthCharacter(k)
    ...
    index = count the number of 1s in the binary representation of (k-1)
    return character at index from string_of_lowercase_letters
\end{lstlisting}
\end{subfigure}

\begin{subfigure}{0.49\textwidth}
\begin{lstlisting}[language=Go, caption={Direct Translation Result}, label={lst: rq1-type2-py2go-direct}]
func kthCharacter(k int) byte {
	...
	return ascii_lowercase[(k-1)^((k-1)&-((k-1)))-1]
    // Calculating a mask instead of counting 1 in the binary expression.
}

// We leverage the space below to show hallucinated implementation of GPT-4o-mini and DSCoder.
// GPT-4o-mini: 
//     return ascii_lowercase[(k-1)&0x1F] //diff func.
// DSCoder:
//     index := (k - 1).bit_count() //non-existent API
\end{lstlisting}
\end{subfigure}\;
\begin{subfigure}{0.49\textwidth}
\begin{lstlisting}[language=Go, caption={Pseudocode-based Translation Result}, label={lst: rq1-type2-py2go-viapseudocode}]
func kthCharacter(k int) byte {
    ...
    index := countBits(k - 1)
    return stringOfLowercaseLetters[index]
}
func countBits(n int) int {
    count := 0   // Counting bits in the integer as
    for n > 0 {  // the Python API bit_count does.
        count += n & 1
        n >>= 1
    }
    return count
}
\end{lstlisting}
\end{subfigure}
\setlength{\abovecaptionskip}{-2pt}
\caption{An Example of Python-to-Go Translation Involving Non-Existent Equivalent APIs in Go (Phi-4)\protect\footref{fn:rq1-example-footnote}}
\label{fig: rq1-type2}
\end{figure}

\subsubsection{Typical Limitations of Pseudocode-based Code Translation.}
\label{subsubsec:limitations}
We also observed three major limitations that cause errors in pseudocode-based code translation due to issues in the generated pseudocode. 
They stem from LLMs' incorrect understanding of original programs, the semantic loss in information transmission, and the ambiguity of natural language. Enhancements to generate more accurate pseudocode are needed to mitigate these limitations and harness the advantages of pseudocode-based code translation.

\begin{enumerate}[leftmargin=3.5em]
    \item[Limit-1] \textit{Pseudocode generated by LLMs may be incorrect, describing an intent or logic inconsistent with those of the original program and misleading the implementation in the target PL.}
\end{enumerate}

Explicit semantic code translation based on pseudocode introduces a compulsory code understanding step during translation. However, LLMs are not free from erroneous code understanding. An incorrect code intent or logic can conversely disturb code translation. 
For example, translating a straightforward Python expression \code{sum  = sum+nums[i] if nums[i]<=sum else nums[i]} into Java may not necessitate the understanding of the code functionality. An incorrect intent conversely misled translation: Qwen32B hallucinated a code intent of \code{sum = max(sum, sum + nums[i])} for this expression in pseudocode, misleading the generation of the Java program. The issue happened with both {\svp} and {\svpp} strategies, while direct translation ({\sd}) leads to a correct translation.
{This limitation was observed in 66\% of analyzed cases where pseudocode-based translation failed while direct translation succeeded.}

\begin{enumerate}[leftmargin=3.5em]
    \item[Limit-2] \textit{Pseudocode generated by LLMs may miss information essential to the reproduction of the complete functionality of the original program.}
\end{enumerate}

LLM-generated pseudocode may also miss certain essential information for implementing the complete functionality of the original program. For example, the data type of variables is often omitted in the LLM-generated pseudocode without any description of precision concerns, leaving operations sensitive to data type (e.g., requiring certain precision) hard to reproduce. Another typical missing information for reproducing code semantics is the logic of complicated algorithms. From a brief description of logic without detailed implementations, LLMs may not accurately reproduce all algorithms. The concise description may also miss customized operations in algorithms, leaving the generated code degenerate into a general implementation lacking the expected features.
{This limitation was observed in 27\% of analyzed cases where direct translation outperformed pseudocode-based translation.}

\begin{enumerate}[leftmargin=3.5em]
    \item[Limit-3] \textit{Natural language description in pseudocode may induce ambiguity in elaboration of code intent and logic.}
\end{enumerate}

Although natural language can depict PL-agnostic code intent and logic, its nature of ambiguity may introduce noise and mislead the code generation step. 
{A typical ambiguity is about the descriptions of loop boundaries using ``to''/``downto'', which cannot clarify the inclusion of border values. 
For example, Qwen32B generated a pseudocode with \code{for i from 0 to length of nums} for the original Rust loop \code{for i in 0..nums.len()}, where the upper border value is excluded. 
When implementing the pseudocode in Go, Qwen32B generated a loop \code{for i:=0; i<=len(nums) \dots} with the upper border value included, resulting in a semantic drift. 
Another common issue is the unclear precedence among arithmetic operations described by natural language descriptions.
For example, Qwen32B generated a pseudocode using \code{a + integer division of (k - 1) by a - 1} to describe the original Java expression \code{a+(k-1)/a-1}. The unclear precedence of addition and subtraction misled Qwen32B to implement \code{a+Math.floor((k-1)/(a-1))} in JavaScript, which drifts from original semantics.
}
Including original programs (i.e., using {\svpp} strategy) cannot mitigate the disturbance caused by the ambiguity as well. In comparison, a direct translation from the original program does not suffer from such ambiguity-induced issues.
{This limitation was observed in 17\% of analyzed cases where direct translation outperformed pseudocode-based translation.}

\subsubsection{Research Opportunities.}

To leverage the advantages of pseudocode-based code translation observed in the previous three RQs and its potential revealed in RQ4, we suggest four directions for future enhancement based on our findings. These target the generation and validation of pseudocode, as well as hybrid approaches to combine the strengths of direct and transitive translation.

\begin{enumerate}[leftmargin=4.4em]
    \item[Chance-1] \textit{Refining pseudocode generation through more fine-grained rules that are tailored for code translation tasks.}
\end{enumerate}

In this study, we rely on generic instructions instructed by \citet{pseudoeval}. Incorporating specific guidelines may help guide LLMs to preserve essential information and minimize ambiguities. The refinement may help capture essential precision requirements, algorithmic details, and clear control logic like loop boundaries that are frequently omitted in standard pseudocode generation, mitigating the information loss and ambiguity that we observed in Limit-2 and Limit-3.

\begin{enumerate}[leftmargin=4.4em]
    \item[Chance-2] \textit{Designing systematic validation and repair mechanisms for generated pseudocode.}
\end{enumerate}

Effective validation strategies may help identify inconsistencies and semantic losses. For example, round-trip validation \cite{roundtripnlp,roundtripmtmt} may help identify buggy pseudocode by translating the generation result back to the source PL for comparison, and LLMs may infer the intended behavior based on code \cite{cruxevalxcao} and compare pseudocode to it to debug pseudocode. Incorporating such validation steps in the loop would help identify and mitigate failures stemming from low-quality pseudocode suffering from the three limitations.

\begin{enumerate}[leftmargin=4.4em]
    \item[Chance-3] \textit{Automating selection of the translation strategy for each task without dynamic validation.}
\end{enumerate}

In this study, we collect the generation results of both strategies and rely on dynamic test execution to determine the optimal strategy. A lightweight static classifier or heuristic that inspects source code characteristics and pseudocode quality for early selection or rejection of translation strategies could streamline the translation process and reduce computational overhead. In addition, the advancement in LLM-based execution prediction may also facilitate validation of candidate translations without dynamic execution~\cite{cruxevalxcao}.

\begin{enumerate}[leftmargin=4.4em]
    \item[Chance-4] \textit{Combining pseudocode-based and direct translations with mutual information.}
\end{enumerate}

Exploring more seamless combination approaches of the two approaches may also further enhance the resulting code translation accuracy. For example, pseudocode could be leveraged to guide the validation and repair of direct translation results, or vice versa. Such integration may help integrate the fine-grained advantages of both approaches.

\begin{enumerate}[leftmargin=4.4em]
    \item[Chance-5] \textit{Enhancing code translation via reasoning.}
\end{enumerate}

Several advanced LLMs (e.g., DeepSeek-R1) have a built-in mechanism to generate reasoning traces in the response process, which proves helpful in guiding LLMs to solve complicated tasks like mathematical and programming problems \cite{deepseekr1techreport}. However, it remains underexplored how to effectively leverage reasoning traces for code translation. There can be several potential directions.
One idea is to directly use reasoning traces as an intermediate representation to guide code translation, similar to pseudocode. However, reasoning traces are often more verbose and less structured than pseudocode, which may introduce unique challenges. For example, we observed serious overthinking issues when prompting DeepSeek-R1 Distilled Qwen7B to generate reasoning traces before code translation, leading to a decrease in translation accuracy compared to direct and pseudocode-based translation and an increase in computation cost, like issues revealed by existing studies \cite{icml25overthinkingreasoning}.
In addition, the adoption of reasoning during pseudocode generation is also worth exploring to help LLMs better understand and depict original programs. 
In this study, we have employed chain-of-thought prompting \cite{cot} to trigger reasoning traces of the five LLMs studied, aiming to guide LLMs to explicitly put down the abstraction and analysis steps that may help generate correct and concise pseudocode.
Nevertheless, more systematic designs to effectively integrate reasoning into code translation are worth further exploration.

\section{Threats to Validity}
\label{sec:threats}

We discuss three potential threats to the validity of our study and our mitigation methods as follows.

\textbf{Representativeness of Studied Subjects.} 
The first threat to our study is about the representativeness and generalizability of the observations on our study subjects. To mitigate this threat, we followed existing work to investigate code translation across six PLs. These include Python, C++, Java, and JavaScript, which are widely adopted in daily development and studied in existing works \cite{tiobeindex,transcoder-ct, ase23ctstudysjtu, fse24llmtranslation}, as well as the emerging Go and Rust \cite{icse24translationllmsurvey, icse25intertrans}. 
Our study results are expected to guide the adoption of pseudocode-based translation for these popular PLs.

We constructed translation tasks based on various LeetCode problems in LiveCodeBench \cite{livecodebench}. LeetCode problems are widely used in code translation studies \cite{oopslatranspilation,avataracl} and benchmarking LLMs' coding ability \cite{qwen25techreport,dpsv3techreport,livecodebench} since they provide high-quality tests to evaluate candidate solutions and provide easy access to collect multilingual programs. 
Also, they require common engineering and algorithm implementations useful in daily development. Thus, we consider the study results on their solution programs should be meaningful and generalizable to practical development code.

We used five LLMs deployable on our machine as the code translator, including both general and coding LLMs, and both closed-source and open-source LLMs from four families. They rank a varying range (from Top-10 to 68 according to the record on Oct 1, 2025) on the famous BigCodeBench \cite{bigcodebench}. Also, the training data for these LLMs has a cutoff date with relatively little overlap with the timeline of the LeetCode problems to construct translation tasks. Therefore, we consider the results learned with these LLMs should be representable and reliable.

\textbf{Data Contamination.}
A potential threat to our study result is the subject to the data contamination issue, where LLMs perform well because they have learned the ground truth during training, rather than pseudocode-based translation strategies. 
Although our translation tasks built upon LiveCodeBench are much newer than the subjects in the conventional code translation benchmarks as introduced in \Cref{subsec:plandtranstasks}, there is still some overlap between the timeline of LiveCodeBench problems and the studied LLMs' training data. 
To further verify the cause of the observed performance improvement, we re-evaluate LLMs on the solution programs of a clean subset of 114 problems released after the latest knowledge cutoff of the studied LLMs, i.e., June 1, 2024. 
As shown in \Cref{table: threat-contamination}, the performance improvement of different strategies is generally consistent with that observed on all 323 problems. Therefore, we consider the observation based on the complete set to be meaningful and provide more statistically meaningful observations with more subjects.

\begin{table}[t]
    \caption{{Pass@10 Rates on All 323 Problems and 114 Newer Problems After Cut-off Date (2024-06-01)}\label{table:
    threat-contamination}}
    \footnotesize
    \setlength{\tabcolsep}{2.4pt}
    \begin{tabular}{l|cc@{\hskip 1pt}cc@{\hskip 1pt}c|cc@{\hskip 1pt}cc@{\hskip 1pt}c}
        \toprule \textbf{Strategy}                 & \textbf{LLM}                                                            & \multicolumn{2}{c}{\textbf{All 323 Probs}} & \multicolumn{2}{c|}{\textbf{\begin{tabular}[c]{@{}c@{}}114 Newer Probs\end{tabular}}} & \textbf{LLM}    & \multicolumn{2}{c}{\textbf{All 323 Probs}} & \multicolumn{2}{c}{\textbf{\begin{tabular}[c]{@{}c@{}}114 Newer Probs\end{tabular}}} \\
        \midrule {\sd}\;{\scriptsize (Direct Translation)} & \multirow{5}{*}{\begin{tabular}[c]{@{}c@{}}Qwen\\ 32B\end{tabular}}     & 0.8777                                     &                                                                                       & 0.8604          &                                        & \multirow{5}{*}{\begin{tabular}[c]{@{}c@{}}Qwen\\ 7B\end{tabular}}                             & 0.7971          &                                   & 0.7623          &                                   \\
        {\svp}\;{\scriptsize (viaPseudocode)}       &                                                                         & 0.8703                                     & {\footnotesize(-0.84\%)}                                                              & 0.8424          & {\footnotesize(-2.09\%)}               &                                                                                                & 0.6193          & {\footnotesize(-22.31\%)}         & 0.5320          & {\footnotesize(-30.21\%)}         \\
        {\svpp}\;{\scriptsize (viaPseudocode w/ ctx.)} &                                                                         & 0.9023                                     & {\footnotesize(+2.80\%)}                                                              & 0.8848          & {\footnotesize(+2.84\%)}               &                                                                                                & 0.7954          & {\footnotesize(-0.21\%)}          & 0.7435          & {\footnotesize(-2.47\%)}          \\
        {\sdvp}\;{\scriptsize (Hybrid[DT\&vP])}     &                                                                         & \textbf{0.9325}                            & \textbf{{\footnotesize(+6.24\%)}}                                                     & \textbf{0.9199} & \textbf{{\footnotesize(+6.92\%)}}      &                                                                                                & 0.8363          & {\footnotesize(+4.92\%)}          & 0.7956          & {\footnotesize(+4.37\%)}          \\
        {\sdvpp}\;{\scriptsize (Hybrid[DT\&vPw/c.])} &                                                                         & 0.9292                                     & {\footnotesize(+5.87\%)}                                                              & 0.9170          & {\footnotesize(+6.58\%)}               &                                                                                                & \textbf{0.8523} & \textbf{{\footnotesize(+6.93\%)}} & \textbf{0.8151} & \textbf{{\footnotesize(+6.93\%)}} \\
        \midrule {\sd}\;{\scriptsize (Direct Translation)} & \multirow{5}{*}{Phi-4}                                                  & 0.8363                                     &                                                                                       & 0.8146          &                                        & \multirow{5}{*}{\begin{tabular}[c]{@{}c@{}}DS\\ Coder\end{tabular}}                            & 0.7934          &                                   & 0.7672          &                                   \\
        {\svp}\;{\scriptsize (viaPseudocode)}          &                                                                         & 0.7839                                     & {\footnotesize(-6.27\%)}                                                              & 0.7702          & {\footnotesize(-5.45\%)}               &                                                                                                & 0.7446          & {\footnotesize(-5.23\%)}          & 0.7185          & {\footnotesize(-6.35\%)}          \\
        {\svpp}\;{\scriptsize (viaPseudocode w/ ctx.)} &                                                                         & 0.8398                                     & {\footnotesize(+0.42\%)}                                                              & 0.8210          & {\footnotesize(+0.79\%)}               &                                                                                                & 0.8016          & {\footnotesize(+3.30\%)}          & 0.7792          & {\footnotesize(+1.56\%)}          \\
        {\sdvp}\;{\scriptsize (Hybrid[DT\&vP])}     &                                                                         & \textbf{0.8926}                            & \textbf{{\footnotesize(+6.73\%)}}                                                     & \textbf{0.8773} & \textbf{{\footnotesize(+7.70\%)}}      & \textbf{}                                                                                      & 0.8587          & {\footnotesize(+8.23\%)}          & 0.8330          & {\footnotesize(+8.58\%)}          \\
        {\sdvpp}\;{\scriptsize (Hybrid[DT\&vPw/c.])} &                                                                         & 0.8886                                     & {\footnotesize(+6.25\%)}                                                              & 0.8735          & {\footnotesize(+7.23\%)}               &                                                                                                & \textbf{0.8616} & \textbf{{\footnotesize(+8.60\%)}} & \textbf{0.8359} & \textbf{{\footnotesize(+8.95\%)}} \\
        \midrule {\sd}\;{\scriptsize (Direct Translation)} & \multirow{5}{*}{\begin{tabular}[c]{@{}c@{}}GPT-4o\\ -mini\end{tabular}} & 0.8539                                     &                                                                                       & 0.8363          &                                        & \multirow{5}{*}{\textit{(Avg)}}                                                                & 0.8317          &                                   & 0.8082          &                                   \\
        {\svp}\;{\scriptsize (viaPseudocode)}          &                                                                         & 0.7409                                     & {\footnotesize(-13.23\%)}                                                             & 0.7155          & {\footnotesize(-14.44\%)}              &                                                                                                & 0.7518          & {\footnotesize(-9.60\%)}          & 0.7157          & {\footnotesize(-11.44\%)}         \\
        {\svpp}\;{\scriptsize (viaPseudocode w/ ctx.)} &                                                                         & 0.8590                                     & {\footnotesize(+0.60\%)}                                                              & 0.8520          & {\footnotesize(+1.88\%)}               &                                                                                                & 0.8396          & {\footnotesize(+0.95\%)}          & 0.8161          & {\footnotesize(+0.98\%)}          \\
        {\sdvp}\;{\scriptsize (Hybrid[DT\&vP])}     &                                                                         & 0.9031                                     & {\footnotesize(+5.76\%)}                                                              & 0.8857          & {\footnotesize(+5.91\%)}               &                                                                                                & 0.8846          & {\footnotesize(+6.37\%)}          & 0.8623          & {\footnotesize(+6.70\%)}          \\
        {\sdvpp}\;{\scriptsize (Hybrid[DT\&vPw/c.])} &                                                                         & \textbf{0.9093}                            & \textbf{{\footnotesize(+6.49\%)}}                                                     & \textbf{0.8962} & \textbf{{\footnotesize(+7.16\%)}}      & \textbf{}                                                                                      & \textbf{0.8882} & \textbf{{\footnotesize(+6.80\%)}} & \textbf{0.8675} & \textbf{{\footnotesize(+7.35\%)}} \\
        \bottomrule
    \end{tabular}
\end{table}

\textbf{Representativeness of Studied Translation Strategies.}
The representativeness of the studied translation strategies and prompts is another potential threat to our study, affecting the meaningfulness of our observations. To mitigate this threat, we studied five code translation strategies, including one direct translation strategy commonly used in LLM-based code translation studies and practices \cite{icse24translationllmsurvey,fse24llmtranslation}, as well as two pseudocode-based translation strategies and two hybrid translation strategies. The pseudocode-based strategies emulate the human practices of semantic translation on natural languages \cite{newmark1981semantictranslation} and are implemented in a transitive translation manner following \cite{icse25intertrans}. The hybrid strategies are based on straightforward result combination to explore the helpfulness of the combined advantages of strategies, considering the feasibility of selecting multiple translation candidates based on original programs \cite{icse25intertrans}. We carefully designed the prompts following existing studies of direct code translation \cite{fse24llmtranslation,icse24translationllmsurvey} and pseudocode-based code generation \cite{pseudoeval}. Thus, they can reflect the typical usage of these strategies and prompts and lead to meaningful observations.

\section{Related Work}
\label{sec:relatedwork}

\subsection{Automated Code Translation}

To facilitate development activities relying on code translation, various automated code translation methods have emerged in the past decades \cite{cxcodetranssurvey,ase23ctstudysjtu}. \citet{fse13statisticalcodetranslation} pioneered in automated code translation via statistical machine translation. To improve translation accuracy by operating directly on ASTs or program graphs, researchers designed tree- and graph-based neural models \cite{Tree-to-tree, graphcodebert}. Unsupervised and self-supervised model training approaches were later proposed to mitigate the scarcity of parallel code corpora \cite{TransCoder, transcoder-ct, multiplatform-1}.

Given the impressive capabilities of LLMs in various coding tasks \cite{llm4sesurveynju}, recent works have explored LLMs for code translation. \citet{fse24llmtranslation} demonstrated that LLMs can serve as effective code translators. \citet{icse24translationllmsurvey} summarized the common errors made by LLMs in code translation, aiming to inspire future enhancement of LLM-based code translation. \citet{fse25alphatrans} leveraged LLMs for project-level code translation. These works mainly focused on direct translation from source PL to target PL, where LLMs are fed with the original program and output the translated program. Recently, \citet{icse25intertrans} found that a one-step direct translation from source PL to target PL may not leverage the full potential of LLMs due to the varying gap between source and target PLs. They proposed transitive direct translations with an intermediary PL as a bridge.

Existing automated code translation approaches mainly focus on code-to-code transformation. Differently, we study the effectiveness of emulating semantic translation for code via pseudocode, where LLMs first interpret the semantics of original programs with pseudocode and then produce translated programs. 
We identify the complementary role of pseudocode-based translation for conventional direct translation, as well as its helpfulness in decoupling task burdens and bridging the differences between the features and APIs in varying PL pairs. The findings may guide practitioners to obtain accurate LLM-generated code translations in a new manner through abstraction.

In addition, some works attempted to debug the erroneous code translations, e.g., by locating code snippets leading to inconsistent behaviors between the original and translated programs to guide repair \cite{transmap} and conducting certain types of repair based on fixed templates \cite{BatFix}. These methods mainly debug minor mistakes in translation results. They are generally orthogonal to our pseudocode-based code translation approach, which aims to reduce the occurrence of errors in the translation results and may also facilitate patch generation. Our methods can be combined with the debugging methods to further enhance the accuracy of LLM-generated code translations.

\subsection{Applications of Pseudocode}

Existing works exploit pseudocode for diverse code-related tasks. A popular task is to generate concrete programs from pseudocode. \citet{Dir17} designed an XML-based method to help novice programmers learn PLs based on programs generated from pseudocode. \citet{Kul19} proposed a search-based method and \citet{Ach22} applied Seq2Seq neural models to enhance the code generation performance. Pseudocode is also used as an intermediate representation to enhance multilingual code generation accuracy \cite{unicoder}. Recently, \citet{pseudoeval} annotated pseudocode for LeetCode problems to isolate the evaluation of problem-solving and language-coding abilities of LLMs in code generation. They also reveal that LLMs can generate programs in different PLs based on pseudocode and synthesize pseudocode for concrete programs, which motivates our study.

In addition to code generation, pseudocode has also proven useful in cross-PL code retrieval by embedding structured pseudocode representations \cite{iclr25pcoderetrieval}, in binary function similarity and vulnerability search by leveraging decompiled pseudocode to obtain platform-agnostic semantic representations \cite{cybersec23pcodebinaryanalysis}, and in generating descriptive summaries for stripped binaries using pseudocode extracted by decompilers as an expert-guided signal \cite{emnlp23pcodesummarization}. All these applications leveraged the advantage of pseudocode in succinctly capturing the intent and logic of programs to facilitate code comprehension and semantic representation \cite{ase15pseudocodegensmt}.

This work studies the helpfulness of pseudocode for another common software development task, i.e., code translation, where pseudocode serves as a semantic interpretation for emulating semantic translation. 
The extensive investigation of pseudocode's effectiveness in code translation yields various findings, which may guide practitioners to obtain accurate LLM-generated code translations based on pseudocode. We also identified limitations of pseudocode generated by LLMs, which are expected to shed light on the focus of validation and repair for automatically generated pseudocode, helping harness the potential of pseudocode for downstream tasks like code translation.

\section{Conclusion}
\label{sec:conclusion}

In this work, we empirically study the effectiveness of pseudocode-based code translation with LLMs, which explicitly emulates the human practice of semantic translation. 
By investigating the performance of five popular LLMs with pseudocode-based code translation on 9,690 translation tasks across six popular PLs, we reveal that pseudocode-based translation can effectively complement the widely adopted direct translation approach for various pairs of source and target PLs. We also identify the highly benefited PL pairs and demonstrate the effectiveness of pseudocode as a general and effective intermediary, as well as the further potential of translation based on higher-quality pseudocode and the bottleneck in both code understanding and generation. 
Our case studies further identify the advantages and limitations of pseudocode-based code translation.
Based on these findings, we suggest the hybrid use of pseudocode-based and direct translation approaches to enhance code translation accuracy. We also discuss future research directions to further unleash the potential of pseudocode in code translation.

\section*{Acknowledgments}

We would like to thank the anonymous reviewers for their insightful comments.
This work is partially supported by the National Natural Science Foundation of China (Grant No. 92582201), the Hong Kong SAR Research Grant Council/General Research Fund (Ref No. 16210725), and the Hong Kong SAR Research Grant Council/Theme-based Research Scheme (Ref No. T41-517/25-N).

\clearpage

\bibliographystyle{ACM-Reference-Format}
\bibliography{references}

\end{document}